# Crystal Instabilities and Elastic Responses of Metals under Extreme Strain Rates

Kun Wang[a], Jun Chen[a,b*], Wenjun Zhu[c], Meizhen Xiang[a]

[a] *Laboratory of Computational Physics, Institute of Applied Physics and Computational Mathematics, Beijing 100088, PR China*
[b] *Center for Applied Physics and Technology, Peking University, Beijing 100071, China*
[c] *National Key Laboratory of Shock Wave and Detonation Physics, Institute of Fluid Physics, Mianyang 621900, China*

## Abstract

Elastic behaviors of crystals often exhibit complex nonlinearities under dynamic loadings. And recent investigations on some metals under ramp compressions show that singularities may arise in the elastic wave and lead to formations of elastic shock waves at early stages of the ramp. It is found that the singularities are due to elastic instabilities triggered by strain as well as strain gradients emerged in the ramped crystals. However, traditional crystal instabilities are discussed within scopes of strain-based elasticity theory. Despite of some progresses in investigating the roles of the higher-order strain gradients on elastic stabilities of solids, the physical nature on the higher-order elastic instabilities of crystals, especially under extreme strain rates, is still a mystery. In this work, a generalized elastic instability criterion for infinite crystals is consistently established at both continuum and atom level under frameworks of a higher-order phenomenological theory. The established criterion could consistently reproduce the well-known strain-based lattice instability criteria, such as modified Born criterion, Λ-criterion, as well as a higher-order one proposed by Bardenhagen et al. Our results show that modified Born criterion is not as precise as the Λ-criterion under heterogeneous stress states. Different from the higher-order criterion, contributions from the third order gradients of displacements are considered so that the well-known "sign" paradox in the first strain-gradient theory could be reproduced. According to microscopic comprehensions on the higher-order phenomenological theory, the "sign" paradox is well clarified. Finally, the established criterion is employed to investigate the elastic stabilities of single crystalline copper and aluminum. The obtained results could well explain the singularities of elastic responses of the two metals under ramp compressions.

## I. Introduction

Response of solids to extreme strain rates is of particular interest to many industrial, transportation, and defense applications in shocks and impacts [1]. Present fundamental equations of the nonlinear dynamics are based on nonlinear elasticity theory, characterized by higher order elastic constants [2,3]. Under the theoretical framework, the nonlinear elastic response of shocked single crystals along various crystallographic orientations could be well understood [4,5]. However, recent investigations [6,7] on the shock response of iron under ramp compressions uncover that singularities may emerge in elastic waves before plasticity or phase transition begins, which is unexceptional under present nonlinear elasticity theory. Under extreme strain rates, comparable to atom thermal vibrational frequencies, deformation mechanisms of materials are in fact governed by lattice instabilities rather than thermal activations. Indeed, it is found that the singularities could be well explained by strain-gradient induced elastic instabilities. In fact, this phenomenon is widely observed in metals under ramp compressions through the non-equilibrium molecular dynamics simulations, such as aluminum and copper as illustrated in present work, but their physical nature is not thoroughly understood.

---

[*] Corresponding authors. Email: jun_chen@iapcm.ac.cn (J. Chen).

It is known that size and microstructure effects cannot be averaged out in macroscopic response when the relative macroscopic dimension is comparable to the characteristic size of the microstructures. Such situation is often encountered in nano-materials. The same could also occur in shocked materials where width of wave front due to the shock is comparable to the microstructure (lattice constants or microscopic characteristic length). These effects could be successfully captured in continuum theories that involve a material length scale, such as strain gradient related theories [8-10]. And the role of higher-order gradients on the stabilities of solids has been investigated by Bardenhagen et al [11,12] through consistently deriving higher-order displacement gradient models from the microscale one. Their studies show that the simplest possible mode that takes the size effects into account is the second order displacement gradient one, corresponding to the first strain gradient theory. Despite of the extensive literatures on strain-gradient elasticity, the theory still appears to be phenomenological because strain-gradient elastic (EGE) constants are unknown and thus merely serve as fitting parameters for practical purpose. To overcome these drawbacks, a lattice dynamic approach is established to determine the SGE constants [8,13-15]. Key point of the approach is to numerically fitting phonon dispersion relations along certain high symmetry directions in order to acquire the requisite elastic constants. However, great cautions should be taken when performing the fitting procedures. For example, the fit should be carried out starting from k-vectors in the vicinity of zero to the one where dispersive effects just start to kick in [13]. Fitting at k-vector, corresponding to regions where frequencies are very high and dispersive effects are large, will results in spurious estimations of the elastic constants. Alternatively, a statistical mechanics approach, which relates the SGE constants to atomic displacement correlations in a molecular dynamics (MD) ensemble, is proposed by Maranganti et al [13]. This approach could be also applied for estimating the elastic constants of non-crystalline systems. To acquire the elastic constants with high precision, a large simulation cell should be employed in the MD simulations. Both approaches adopt a "dynamic" or "indirect" (statistical) way to acquire a proper estimation of the elastic constants. Most recently, Admal et al [16] propose a "direct" approach at atom level to obtain the SGE constants. Their derivations are based on condition of energetic equivalence between continuum and atomistic representations of crystals when kinetics of the latter is governed by the Cauchy-Born rule. Their starting point is based on a polynomial expansion of deformation map which is conceptually a continuum approach.

It has been well recognized that the sign of the second displacement gradient elastic constants in the first strain gradient theory is opposite with the one of curvature of the dispersion curve[17,18], known "sign" paradox. When considering the higher-order elastic stabilities in real crystals, the "sign" paradox would emerge and result in two different stability conditions, but previous works on the solid stabilities neither explain which stability condition is physically acceptable nor clarify their relationships with commonly used crystal instability criteria. For such reason, a generalized elastic instability criterion is established, which could encompass both the traditional strain-based instability criteria and the higher-order instability condition under a unified theoretical framework at continuum level. Some new insights into both the strain-based instability criteria and the higher-order one are provided. Further, the generalized elastic instability criterion is comprehended at atom level through both static and dynamic approaches. Consequently, the well-known "sign" paradox of the SGE constants appears, which is explained within scopes of present work. In addition, the generalized elastic instability criterion is employed to investigate stabilities of two metals, i.e., single crystalline copper and aluminum, under uniaxial compressions and the obtained results are compared with ones from direct non-equilibrium molecular dynamics simulations.

The remainder of present work is organized as follows. In Part II, higher-order elastic instability criterion is established within frameworks of a high-order phenomenological continuum theory, which enable us to consistently reproduce many well-known results on crystal instabilities. Then, two typical microscopic comprehensions on the established instability criteria are provided in Part III. Finally, in Part IV and V, the generalized elastic instability criterion is employed to investigate elastic instabilities of single crystalline copper and aluminum, and non-equilibrium molecular dynamic simulations are conducted to confirm our conclusions.

## II.  Higher Order Elastic Instabilities

In this part, we firstly derive motion equations and the corresponding boundary conditions based on higher-order continuum elastic theories which are pioneered by Mindlin [10] and DiVincenzo [8] respectively via static and dynamic approaches, and later developed by others [11,19,20], to name a few. Then higher-order elastic instability criterion (HEIC) for infinite crystals is established. And links between the established HEIC and the traditional strain-based mechanical instability criteria (SMIC) are clarified. As a result, some new insights into these constantly used SMIC are obtained. By convention, letter in bold represent vector or tensor, otherwise, is a scalar. If not specified, English alphabet in the subscript denotes Cartesian index of the corresponding vector or tensor, and Greek alphabet in the superscript stands for which atoms the corresponding quality is defined on. Einstein summation convention over repeated subscript is employed throughout this work. For brevity, gradient of a tensor $\mathbf{A}$ is expressed by $\mathbf{A}\nabla = \partial A_{i,\dots,j}/X_k$, and divergence of $\mathbf{A}$ is $\mathbf{A} \cdot \nabla = \partial A_{i,\dots,j}/X_j$. The first and second order time derivative of a vector $\mathbf{v}$ is denoted by $\dot{\mathbf{v}}$ and $\ddot{\mathbf{v}}$, respectively.

### A. Higher-order phenomenological continuum theory

Kinetics of deformations in traditional continuum mechanics are described by deformation gradient $\mathbf{F}$, a two-point tensor, which relates initial position $\mathbf{X}$ in a material volume $\mathfrak{B}_0$ to final position $\mathbf{x}$ in spatial (deformed) volume $\mathfrak{B}_1$ by $\mathbf{F} = \mathbf{x}\nabla$, where $\nabla$ is the material gradient operator. In higher-order continuum theory, the second-order deformation gradient $\mathbf{G} = \mathbf{F}\nabla$, a rank-three tensor, is also introduced in addition. And displacement of arbitrary material point is $\mathbf{u} = \mathbf{x} - \mathbf{X}$, which relates to $\mathbf{F}$ and $\mathbf{G}$ by

$$\mathbf{F} = \mathbf{u}\nabla - \mathbf{I}, \quad \mathbf{G} = \mathbf{u}\nabla\nabla, \tag{1}$$

where $\mathbf{I}$ is the unit dyadic. Then a phenomenological expression for the Lagrangian density "$\mathcal{L}$" for a solid [8] is

$$\mathcal{L} = \tfrac{1}{2}\rho \dot{\mathbf{u}} \cdot \dot{\mathbf{u}} - V^{tot}(T, \mathbf{F}, \mathbf{G}), \tag{2}$$

where $\rho$ is the mass density of the solid and $T$ stands for temperature at the material point considered. In above expression, the first term represents kinetic energy and the second term is the specific potential energy of the solid which depends on the temperature and the deformation states, represented by gradients of deformation up to the second order. It should be noted that the potential energy relies on time implicitly through the time evolution of $\mathbf{F}$ and $\mathbf{G}$. To work with the higher-order elastic instabilities, only elastic deformation processes are needed to be considered. If the temperature of the surroundings is unchanged during the deformation, i.e., an isothermal process, then the potential energy corresponds to the specific Helmholtz free energy. When an adiabatic process is considered, the potential energy corresponds to the specific internal energy. Deformations of solids under shock compressions belong to the latter case. Since no temperature variations in the elastic deformations, we will drop $T$ in the following derivations. For the variations in the action "$A$" to be stationary with respective to small virtual variations in displacements, we have

$$\delta A = \delta \int \int \mathcal{L} dV dt = \int \int \rho \delta \dot{\mathbf{u}} \cdot \dot{\mathbf{u}} dV \, dt - \int \delta E^{tot} dt = 0, \tag{3}$$

where

$$\delta E^{tot} = \int_{\mathcal{B}_0} \delta V^{tot}(\mathbf{F}, \mathbf{G}) \mathrm{d}V = \int_{\mathcal{B}_0} \delta W(\mathbf{F}, \mathbf{G}) \mathrm{d}V + \delta E^{ext}$$

$$= \int_{\mathcal{B}_0} \left( \frac{\partial W}{\partial \mathbf{F}} : \delta \mathbf{u} \nabla + \frac{\partial W}{\partial \mathbf{G}} \vdots \delta \mathbf{u} \nabla \nabla \right) \mathrm{d}V + \delta E^{ext}$$

$$= \int_{\mathcal{B}_0} \left[ -\delta \mathbf{u} \cdot \left( \frac{\partial W}{\partial \mathbf{F}} \cdot \nabla \right) + \delta \mathbf{u} \cdot \left( \frac{\partial W}{\partial \mathbf{G}} : \nabla \nabla \right) \right] \mathrm{d}V + \int_{\partial \mathcal{B}_0} \delta \mathbf{u} \cdot \left( \frac{\partial W}{\partial \mathbf{F}} - \frac{\partial W}{\partial \mathbf{G}} \cdot \nabla \right) \cdot \mathbf{N} \mathrm{d}S$$

$$+ \int_{\partial \mathcal{B}_0} \delta \mathbf{u} \nabla : \frac{\partial W}{\partial \mathbf{G}} \cdot \mathbf{N} \mathrm{d}S + \delta E^{ext} = 0.$$

$$\tag{4}$$

Variations of external potentials, $\delta E^{ext}$, are also involved in the variations of total potentials ($\delta E^{tot}$). It is worth noting that $\delta \mathbf{u}$ and the surface components of its gradient are not independent since the derivatives of $\delta \mathbf{u}$ along directions tangent to surface $\partial \mathcal{B}_0$ could be determined if all $\delta \mathbf{u}$ at the surface is known. Thereby, the last boundary term in last equation of (4) should be expressed further in terms of independent variables in $\partial \mathcal{B}_0$. Proceeding as by Mindlin [9,10], $\delta \mathbf{u} \nabla$ is resolved into a surface-gradient and a normal gradient:

$$\delta \mathbf{u} \nabla = \delta \mathbf{u} \nabla^s + (\delta \mathbf{u} D) \mathbf{N}. \tag{5}$$

where $(\cdot)\nabla^s = (\cdot)\nabla \cdot (\mathbf{I} - \mathbf{N} \otimes \mathbf{N})$, $(\cdot)D = ((\cdot)\nabla) \cdot \mathbf{N}$ and $\mathbf{N}$ is the unit outward normal to the surface ($\partial \mathcal{B}_0$) of the volume ($\mathcal{B}_0$). For arbitrary vector $\mathbf{v}$ and rank-two tensor $\mathbf{\Phi}$, we have

$$\int_{\partial \mathcal{B}_0} (\mathbf{v} \nabla) : \mathbf{\Phi} \mathrm{d}S = \int_{\partial \mathcal{B}_0} [(\mathbf{v} \nabla^s) : \mathbf{\Phi} + (\mathbf{v} D \mathbf{N}) : \mathbf{\Phi}] \mathrm{d}S$$

$$= \int_{\partial \mathcal{B}_0} (\mathbf{v} \cdot \mathbf{\Phi}) \cdot \nabla^s \mathrm{d}S - \int_{\partial \mathcal{B}_0} \mathbf{v} \cdot (\mathbf{\Phi} \cdot \nabla^s) \mathrm{d}S + \int_{\partial \mathcal{B}_0} (\mathbf{v} D \mathbf{N}) : \mathbf{\Phi} \mathrm{d}S$$

$$= \int_{\partial \mathcal{B}_0} (\mathbf{N} \cdot \nabla^s)(\mathbf{v} \cdot \mathbf{\Phi} \cdot \mathbf{N}) \mathrm{d}S - \int_{\partial \mathcal{B}_0} \mathbf{v} \cdot (\mathbf{\Phi} \cdot \nabla^s) \mathrm{d}S + \int_{\partial \mathcal{B}_0} \mathbf{v} D \cdot \mathbf{\Phi} \cdot \mathbf{N} \mathrm{d}S$$

$$= \int_{\partial \mathcal{B}_0} \mathbf{v} \cdot \mathbf{\Phi} \cdot (\mathbf{N} \otimes \mathbf{N} \cdot \boldsymbol{\nabla}^s - \boldsymbol{\nabla}^s) \mathrm{d}S + \int_{\partial \mathcal{B}_0} \mathbf{v} D \cdot \mathbf{\Phi} \cdot \mathbf{N} \mathrm{d}S$$

$$= \int_{\partial \mathcal{B}_0} \mathbf{v} \cdot \mathbf{\Phi} \cdot \mathbf{L} \mathrm{d}S + \int_{\partial \mathcal{B}_0} \mathbf{v} D \cdot \mathbf{\Phi} \cdot \mathbf{N} \mathrm{d}S, \tag{6}$$

where operator $\mathbf{L}$ is defined by

$$\mathbf{L} \stackrel{\text{def}}{=} \mathbf{N} \otimes \mathbf{N} \cdot \boldsymbol{\nabla}^s - \boldsymbol{\nabla}^s. \tag{7}$$

Divergence theorem for smooth, closed surfaces has been used to obtain the third equation of (6), that is

$$\int_{\partial \mathcal{B}_0} \mathbf{w} \cdot \boldsymbol{\nabla}^s \mathrm{d}S = \int_{\partial \mathcal{B}_0} (\mathbf{N} \cdot \boldsymbol{\nabla}^s)(\mathbf{w} \cdot \mathbf{N}) \mathrm{d}S, \tag{8}$$

where $\mathbf{w}$ is an arbitrary vector. Thereby, Eq. (4) reduces to

$$\delta E^{tot} = \int_{\mathcal{B}_0} \left[ -\delta \mathbf{u} \cdot \left( \frac{\partial W}{\partial \mathbf{F}} \cdot \boldsymbol{\nabla} \right) + \delta \mathbf{u} \cdot \left( \frac{\partial W}{\partial \mathbf{G}} : \boldsymbol{\nabla}\boldsymbol{\nabla} \right) \right] \mathrm{d}V + \int_{\partial \mathcal{B}_0} \delta \mathbf{u} \cdot \left( \frac{\partial W}{\partial \mathbf{F}} - \frac{\partial W}{\partial \mathbf{G}} \cdot \boldsymbol{\nabla} \right) \cdot \mathbf{N} \mathrm{d}S$$

$$+ \int_{\partial \mathcal{B}_0} \delta \mathbf{u} \cdot \frac{\partial W}{\partial \mathbf{G}} \cdot \mathbf{N} \cdot \mathbf{L} \mathrm{d}S + \int_{\partial \mathcal{B}_0} \delta \mathbf{u} D \cdot \frac{\partial W}{\partial \mathbf{G}} \cdot \mathbf{N} \cdot \mathbf{N} \mathrm{d}S + \delta E^{ext}$$

$$= \int_{\mathcal{B}_0} \delta \mathbf{u} \cdot \left( -\frac{\partial W}{\partial \mathbf{F}} + \frac{\partial W}{\partial \mathbf{G}} \cdot \boldsymbol{\nabla} \right) \cdot \boldsymbol{\nabla} \mathrm{d}V + \int_{\partial \mathcal{B}_0} \delta \mathbf{u} \cdot \left[ \left( \frac{\partial W}{\partial \mathbf{F}} - \frac{\partial W}{\partial \mathbf{G}} \cdot \boldsymbol{\nabla} \right) \cdot \mathbf{N} + \frac{\partial W}{\partial \mathbf{G}} \cdot \mathbf{N} \cdot \mathbf{L} \right] \mathrm{d}S$$

$$+ \int_{\partial \mathcal{B}_0} \delta \mathbf{u} D \cdot \frac{\partial W}{\partial \mathbf{G}} : \mathbf{N} \otimes \mathbf{N} \mathrm{d}S + \delta E^{ext} = 0. \tag{9}$$

According to strain-gradient theory developed in references[10,20], the variation of the external potential could be expressed into three parts, i.e.,

$$\delta E^{ext} = -\int_{\mathcal{B}_0} \delta \mathbf{u} \cdot \mathbf{b_0} \mathrm{d}V - \int_{\partial \mathcal{B}_0} \delta \mathbf{u} \cdot \mathbf{t}_0^p \mathrm{d}S - \int_{\partial \mathcal{B}_0} \delta \mathbf{u} D \cdot \mathbf{t}_0^Q \mathrm{d}S, \tag{10}$$

where $\mathbf{b_0}$ is body force density, $\mathbf{t}_0^p$ is nominal surface traction and $\mathbf{t}_0^Q$ is the second-order stress traction. The first Piola-Kirchhoff stress tensor and the second-order stress are defined by

$$\mathbf{P} \doteq \frac{\partial W}{\partial \mathbf{F}} \text{ and } \mathbf{Q} \doteq \frac{\partial W}{\partial \mathbf{G}}, \tag{11}$$

respectively. Through replacing $\delta E^{ext}$ in Eq. (10) with (9) and substituting the resulting expression for $\delta E^{tot}$ into Eq. (3), we can get

$$\delta A = \int \int_{\mathcal{B}_0} \{ \rho \delta \dot{\mathbf{u}} \cdot \dot{\mathbf{u}} - \delta \mathbf{u} \cdot (-\mathbf{P} + \mathbf{Q} \cdot \boldsymbol{\nabla}) \cdot \boldsymbol{\nabla} + \delta \mathbf{u} \cdot \mathbf{b_0} \} \mathrm{d}V \, dt$$

$$- \int \int_{\partial \mathcal{B}_0} \delta \mathbf{u} \cdot [(\mathbf{P} - \mathbf{Q} \cdot \boldsymbol{\nabla}) \cdot \mathbf{N} + \mathbf{Q} \cdot \mathbf{N} \cdot \mathbf{L} - \mathbf{t}_0^p] \mathrm{d}S \, dt$$

$$- \int \int_{\partial \mathcal{B}_0} \delta \mathbf{u} D \cdot (\mathbf{Q} : \mathbf{N} \otimes \mathbf{N} - \mathbf{t}_0^Q) \mathrm{d}S \, dt = 0. \tag{12}$$

After integrating the time derivative term by parts, the above expression reduces to terms involving the first-order variations of independent variables each. To make the finial expression to be zero for arbitrary $\delta\mathbf{u}$ and $\delta\mathbf{u}D$, the coefficients for these independent variables must equal to zero. After some rearrangements, higher-order motion equations and Neumann-type boundary conditions are obtained as follows:

$$\rho\ddot{\mathbf{u}} = (\mathbf{P} - \mathbf{Q} \cdot \nabla) \cdot \nabla + \mathbf{b_0} \quad \text{in } \mathfrak{B}_0, \tag{13}$$

$$(\mathbf{P} - \mathbf{Q} \cdot \nabla) \cdot \mathbf{N} + \mathbf{Q} \cdot \mathbf{N} \cdot \mathbf{L} = \mathbf{t}_0^p \quad \text{on } \partial\mathfrak{B}_0, \tag{14}$$

$$\mathbf{Q} : \mathbf{N} \otimes \mathbf{N} = \mathbf{t}_0^Q \quad \text{on } \partial\mathfrak{B}_0. \tag{15}$$

For static cases where time derivative terms are vanished, this result reduces to higher-order equilibrium equations and boundary conditions derived by Sunyk and Steinmann[19].

## B. Generalized Elastic Instability Criterion

Elastic instabilities for materials in the internal of the volume $\mathfrak{B}_0$ are ready to be investigated using the higher-order motion equations (See Eq. 13). To this end, linearized elastic constitutive relations are employed, that is

$$\mathbf{P} = \mathbb{L} : \mathbf{F} + {}^1\mathbb{H} : \mathbf{G} + \mathbf{G} : {}^2\mathbb{H} + {}^1\mathbb{E} \vdots \mathbf{H} + \mathbf{H} \vdots {}^2\mathbb{E}, \quad \mathbf{Q} = \mathbb{M} : \mathbf{G} + {}^2\mathbb{H} : \mathbf{F} + \mathbf{F} : {}^1\mathbb{H}, \tag{16}$$

where elastic constants in above relations are defined by

$$\mathbb{L} \doteq \frac{\partial \mathbf{P}}{\partial \mathbf{F}} = \frac{\partial^2 W}{\partial \mathbf{F} \partial \mathbf{F}}, \quad \mathbb{M} \doteq \frac{\partial \mathbf{Q}}{\partial \mathbf{G}} = \frac{\partial^2 W}{\partial \mathbf{G} \partial \mathbf{G}}, \quad {}^1\mathbb{H}_{iJlMN} + {}^2\mathbb{H}_{lMNiJ} \doteq \frac{\partial P_{iJ}}{\partial G_{lMN}} = \frac{\partial^2 W}{\partial F_{iJ} \partial G_{lMN}},$$

$${}^2\mathbb{H}_{iJKlM} + {}^1\mathbb{H}_{lMiJK} \doteq \frac{\partial Q_{iJK}}{\partial F_{lM}} = \frac{\partial^2 W}{\partial G_{iJK} \partial F_{lM}}, \quad {}^1\mathbb{E}_{iJlMNO} + {}^2\mathbb{E}_{lMNOiJ} \doteq \frac{\partial P_{iJ}}{\partial H_{lMNO}} = \frac{\partial^2 W}{\partial F_{iJ} \partial H_{lMNO}},$$

$$\tag{17}$$

where $\mathbf{H} = \mathbf{u}\nabla\nabla\nabla$. It will be clear later in present work that terms in (16), containing $\mathbf{H}$, can also contribute to the instabilities of crystals under the first strain-gradient elasticity theory. As can be seen later in this section, it is more convenient to work with $\mathbb{H}$, $\mathbb{H}^{\ddagger}$ and $\mathbb{E}$ than ${}^1\mathbb{H}$, ${}^2\mathbb{H}$ ${}^1\mathbb{E}$ and ${}^2\mathbb{E}$, where

$$\mathbb{H}_{iJlMN} = {}^1\mathbb{H}_{iJlMN} + {}^2\mathbb{H}_{lMNiJ}, \quad \mathbb{H}^{\ddagger}_{iJKlM} = {}^2\mathbb{H}_{iJKlM} + {}^1\mathbb{H}_{lMiJK}, \tag{18}$$

$$\mathbb{E}_{iJlMNO} = {}^1\mathbb{E}_{iJlMNO} + {}^2\mathbb{E}_{lMNOiJ}. \tag{19}$$

An alternative definition for $\mathbb{H}$, $\mathbb{H}^{\ddagger}$ and $\mathbb{E}$ are

$$\mathbb{H} = \frac{\partial^2 W}{\partial \mathbf{F} \partial \mathbf{G}}, \quad \mathbb{H}^{\ddagger} = \frac{\partial^2 W}{\partial \mathbf{G} \partial \mathbf{F}} \text{ and } \mathbb{E} = \frac{\partial^2 W}{\partial \mathbf{F} \partial \mathbf{H}}. \tag{20}$$

These elastic constitutive relations are the higher-order forms of the generalized Hook's law, which will be further justified from lattice model at atom level in the second section of part III. After substituting (16) into Eq. (13) in absent of the body force, we obtain

$$\rho\ddot{u}_i = \mathbb{L}_{iJlM} u_{l,MJ} + {}^1\mathbb{H}_{iJlMN} u_{l,MNJ} + {}^2\mathbb{H}_{lMNiJ} u_{l,MNJ} + {}^1\mathbb{E}_{iJlMNO} u_{l,MNOJ} + {}^2\mathbb{E}_{lMNOiJ} u_{l,MNOJ}$$
$$- \mathbb{M}_{iJKlMN} u_{l,MNKJ} - {}^2\mathbb{H}_{iJKlM} u_{l,MKJ} - {}^1\mathbb{H}_{lMiJK} u_{l,MKJ}$$

$$\tag{21}$$

in its component form, where $(\cdot)_{,N} = \partial(\cdot)/\partial X_N$, $(\cdot)_{,MN} = \partial^2(\cdot)/\partial X_M \partial X_N$ and so on. A plane-wave solution for Eq. (21) takes the form of $\mathbf{u}(\mathbf{X}, t) = \mathbf{u}^0(\mathbf{X})\exp(-i\omega t)$ where $\omega$ represents frequency of wave and $\mathbf{u}^0(\mathbf{X})$ stands for the spatial part of the displacement in $\mathfrak{B}_0$. Imaging that volume $\mathfrak{B}_0$ is taken from an infinite ideal crystal and the continuous displacement field, $\mathbf{u}^0(\mathbf{X})$, corresponds to the atom displacements at lattice points and their interpolations at intervals of these lattice points, then $\mathbf{u}^0(\mathbf{X})$ holds the lattice translational symmetry, i.e., $\mathbf{u}^0(\mathbf{X}) = \mathbf{u}^0(\mathbf{X} + \mathbf{a}^i)$, $\mathbf{a}^i$ ($i = 1,2,3$) are lattice basis vectors of a primitive cell. Thereby, discrete Fourier series expansion for $\mathbf{u}^0(\mathbf{X})$ is

$$\mathbf{u}^0(\mathbf{X}) = \frac{1}{\sqrt{\mathcal{N}}} \sum_{\Xi} \hat{\mathbf{u}}^0(\Xi) \exp(-i\Xi \cdot \mathbf{X}), \tag{22}$$

where $\hat{\mathbf{u}}^0(\Xi)$ is Fourier transformation for $\mathbf{u}^0(\mathbf{X})$ and $\Xi$ is the wavevector with respective to the initial configuration. According to Bloch theorem, $\mathbf{u}^0(\mathbf{X})$ should have a form of $\mathbf{u}_\Xi^0(\mathbf{X})\exp(i\Xi \cdot \mathbf{X})$ where $\mathbf{u}_\Xi^0(\mathbf{X})$ satisfies $\mathbf{u}_\Xi^0(\mathbf{X}) = \mathbf{u}_\Xi^0(\mathbf{X} + \mathbf{a}^i)$. And in solid state physics, the infinite crystal is assumed to consist of many periodic finite crystals, each of which has a dimension of $\mathcal{N}^1 \times \mathcal{N}^2 \times \mathcal{N}^3$ (totally $\mathcal{N}$) primitive cells along three directions, respectively. This means that periodic boundary condition, $\mathbf{u}^0(\mathbf{X}) = \mathbf{u}^0(\mathbf{X} + \mathcal{N}^i \mathbf{a}^i)$, holds for the finite crystal so that

$$\Xi = \frac{h_i}{\mathcal{N}^i} \mathbf{b}^i, \tag{23}$$

where $h_i (i = 1,2,3)$ are integers and $\mathbf{b}^i (i = 1,2,3)$ are reciprocal lattice basis vectors of the crystals which relates the lattice basis vectors by $\mathbf{a}^i \mathbf{b}^j = 2\pi \delta_{ij}$. Because of the symmetries hold by the crystals, only $\Xi$ in the first Brillouin zone of the reciprocal space needs to be considered. Substituting Eq. (22) into Eq. (21), we can obtain

$$\rho \omega_\Xi^2 \hat{u}_i^0(\Xi) = \mathbb{L}_{iJlM} \hat{u}_l^0(\Xi) \Xi_M \Xi_J - i\mathbb{H}_{iJlMN} \hat{u}_l^0(\Xi) \Xi_M \Xi_N \Xi_J - \mathbb{E}_{iJlMNK} \hat{u}_l^0(\Xi) \Xi_M \Xi_N \Xi_K \Xi_J$$
$$+ \mathbb{M}_{iJKlMN} \hat{u}_l^0(\Xi) \Xi_M \Xi_N \Xi_K \Xi_J + i\mathbb{H}_{iJKlM}^\ddagger \hat{u}_l^0(\Xi) \Xi_M \Xi_K \Xi_J = \mathbb{B}_{iJlM} \Xi_M \Xi_J \hat{u}_l^0(\Xi), \tag{24}$$

or

$$\frac{1}{\rho} \hat{u}_i^0(\Xi) \Xi_J \mathbb{B}_{iJlM} \hat{u}_l^0(\Xi) \Xi_M = \omega_\Xi^2 \|\hat{\mathbf{u}}^0(\Xi)\|^2 \geq 0, \tag{25}$$

where $\|\hat{\mathbf{u}}^0(\Xi)\| = \sqrt{\hat{\mathbf{u}}^0(\Xi) \cdot \hat{\mathbf{u}}^0(\Xi)}$ and

$$\mathbb{B}_{iJlM} = \mathbb{L}_{iJlM} - i\left(\mathbb{H}_{iJlMN} \Xi_N - \mathbb{H}_{iJKlM}^\ddagger \Xi_K\right) + \left(\mathbb{M}_{iJKlMN} - \mathbb{E}_{iJlMNK}\right) \Xi_K \Xi_N. \tag{26}$$

Any violation of condition (25) indicates certain unstable vibrational modes appearing in crystals, and thus elastic instability begins. Without losing generalities, let $\|\hat{\mathbf{u}}^0(\Xi)\| = 1$. To proceed, it is convenient to contract the rank-four tansor $\mathbb{B}$ into an 9X9 matrix $\widehat{\mathbb{B}}$ through $\widehat{\mathbb{B}}_{pq} = \mathbb{B}_{IJLM}$ where the $p \leftrightarrow IJ$ (or $q \leftrightarrow LM$) correspondence is

$$1 \leftrightarrow 11, 2 \leftrightarrow 12, 3 \leftrightarrow 13, 4 \leftrightarrow 21, 5 \leftrightarrow 22, 6 \leftrightarrow 23, 7 \leftrightarrow 31, 8 \leftrightarrow 32, 9 \leftrightarrow 33. \tag{27}$$

Then the condition (25) could be expressed as

$$\mathbf{Y}^T \widehat{\mathbb{B}} \mathbf{Y} \geq 0, \tag{28}$$

where $\mathbf{Y}$ is a column vector with components of $Y_p = [\tilde{\mathbf{u}}^0 \otimes \Xi]_{iJ}$. From Eq. (26), it is easy to find that $\widehat{\mathbb{B}}$ is not a symmetric matrix because of existence of the image parts. However, the above inequality is equivalent to its symmetric form. Considering the relation of $\mathbb{H}_{iJlMN} = \mathbb{H}_{lMNiJ}^\ddagger$, the symmetric form of $\mathbb{B}$ is

$$\mathbb{B}_{iJlM}^S = \mathbb{L}_{iJlM} + \left(\mathbb{M}_{iJKlMN} - \mathbb{E}_{iJlMNK}\right) \Xi_K \Xi_N. \tag{29}$$

It should be noted that non-negative minimum eigenvalue of $\widehat{\mathbb{B}}$ is only a sufficient condition of (28). This is because only five out of nine components of $\mathbf{Y}$ is independent. To obtain its equivalent one, let

$$\mathbb{K}_{il}(\Xi) = \mathbb{B}_{iJlM}^S \Xi_M \Xi_J = \left[\mathbb{L}_{iJlM} + \left(\mathbb{M}_{iJKlMN} - \mathbb{E}_{iJlMNK}\right) \Xi_K \Xi_N\right] \Xi_M \Xi_J$$

$$= [\mathbb{L}_{iJlM} + \mathbb{W}_{iJKlMN}\Xi_K\Xi_N]\Xi_M\Xi_J, \tag{30}$$

where

$$\mathbb{W}_{iJKlMN} = \mathbb{M}_{iJKlMN} - \mathbb{E}_{iJlMNK}. \tag{31}$$

the stability condition (28) is equivalent to

$$\hat{u}_i^0(\Xi)\mathbb{K}_{il}(\Xi)\hat{u}_l^0(\Xi) \geq 0, \tag{32}$$

which is the generalized elastic stability criterion. When neglecting the higher-order contribution from $\mathbb{E}$, this result is equivalent to the one given by Bardenhagen et al [11] except for differences in deformation metrics. Later in Part III of present work, we will show that the higher-order contribution provides an important correction for higher-order instability criterion (32). Specially, when the norm of the wavevector to be a unit like Bardenhagen et al, the sufficient condition of (28) is equivalent to the one that the sum of the minimum eigenvalue of $\mathbb{L}^S$ (symmetric part of $\mathbb{L}$) and $\mathbb{W}$ are non-negative. Because the first part of the equivalent sufficient condition is just the traditional strain-based mechanical instability condition (which will be discussed in the next section), for description convenience, hereafter we will refer to the second part as the "sufficient condition". To facilitate derivations in remaining parts of present work, variables defined in current configurations in (32) are redefined by

$$\tilde{u}_I^0(\Xi) \doteq \hat{u}_i^0(\Xi)F_{iI}, \quad \widetilde{\mathbb{K}}_{IL}(\Xi) \doteq \mathbb{K}_{il}F_{Ii}^{-1}F_{Ll}^{-1} = \left(\widetilde{\mathbb{L}}_{IJLM} + \widetilde{\mathbb{W}}_{IJKLMN}\Xi_K\Xi_N\right)\Xi_M\Xi_J, \tag{33}$$

where

$$\widetilde{\mathbb{L}}_{IJLM} \doteq F_{Ii}^{-1}F_{Ll}^{-1}\mathbb{L}_{iJlM}, \quad \widetilde{\mathbb{W}}_{IJKLMN} \doteq F_{Ii}^{-1}F_{Ll}^{-1}\mathbb{W}_{iJKlMN}. \tag{34}$$

It is obvious that the condition of (32) is still valid when replacing the corresponding variables by the ones with a tilde, that is,

$$\widetilde{\mathbf{u}}^0(\Xi)\widetilde{\mathbb{K}}(\Xi)\widetilde{\mathbf{u}}^0(\Xi) \geq 0. \tag{35}$$

Thereby the generalized elastic stability condition requires that the minimum eigenvalue ($\lambda_c^{\widetilde{\mathbb{K}}}$) of $\widetilde{\mathbb{K}}(\Xi)$ for arbitrary $\Xi$ is non-negative. Any violations of this condition indicate that the crystal at the material point is going to be instable. Thus, condition (35) provides a generalized elastic instability criterion (GEIC) for crystals, which is established under linearized theory of the first strain-gradient elasticity.

## C. Linking to traditional strain-based mechanical instability criteria

In present work, we will show the links between the GEIC and frequently used strain-based instability criteria, i.e., modified Born criterion, $\Lambda$-criterion and phonon instability criterion. To make this paper more compact, the links between the GEIC and the phonon instability criterion are discussed in part III B. When higher than the first-order deformation gradients are not present, motion equation (25) reduces to

$$\rho\omega_\Xi^2 = \left(\hat{u}_i^0\Xi_J\right)\mathbb{L}_{iJlM}\left(\hat{u}_l^0\Xi_M\right). \tag{36}$$

Any physically meaningful solutions of the above equation exist if and only if

$$\Lambda(\hat{\mathbf{u}}^0,\Xi) \doteq \left(\hat{u}_i^0\Xi_J\right)\mathbb{L}_{iJlM}\left(\hat{u}_l^0\Xi_M\right) \geq 0 \tag{37}$$

is satisfied for arbitrary $\hat{\mathbf{u}}^0$ and $\Xi$. Since traditional strain-based instability criteria are formulated in terms of Lagrangian strain (**E**), and its first- and second- order work conjugates, links between criteria (37) and the traditional ones could be established via the definitions of the related quantities, that is,

$$\mathbf{E} \doteq \frac{1}{2}(\mathbf{F}^T\mathbf{F} - \mathbf{I}), \tag{38}$$

$$\mathbf{S} \doteq \frac{\partial \widetilde{W}(\mathbf{E})}{\partial \mathbf{E}}, \tag{39}$$

$$\mathbf{C} \doteq \frac{\partial \mathbf{S}(\mathbf{E})}{\partial \mathbf{E}} = \frac{\partial^2 \widetilde{W}(\mathbf{E})}{\partial \mathbf{E} \partial \mathbf{E}}, \tag{40}$$

where $\widetilde{W}(\mathbf{E}, \mathbf{G}) = W(\mathbf{F}, \mathbf{G})$, $\mathbf{S}$ is the second Piola–Kirchhoff stress and $\mathbf{C}$ is the elastic stiffness tensor. Using Eq. (38-40) and (17), the relationship between $\mathbb{L}$ and $\mathbf{C}$ could be obtain, that is,

$$\mathbb{L}_{iJlM} = C_{IJLM} F_{iI} F_{lL} + \delta_{il} S_{JM}, \tag{41}$$

where $\delta_{il}$ is the Kronecker delta. It is convenient to express $\mathbb{L}$ in initial configuration through (34) so that

$$\widetilde{\mathbb{L}}_{IJLM} = C_{IJLM} + F_{Ii}^{-1} F_{Ll}^{-1} \delta_{il} S_{JM}. \tag{42}$$

Considering relation of $S_{JM} = J F_{Jj}^{-1} F_{Mm}^{-1} \sigma_{jm}$ and $C_{IJLM} = J F_{Ii}^{-1} F_{Jj}^{-1} F_{Ll}^{-1} F_{Mm}^{-1} c_{ijlm}$, the above equation could also expressed as

$$\widetilde{\mathbb{L}}_{IJLM} = J F_{Ii}^{-1} F_{Jj}^{-1} F_{Ll}^{-1} F_{Mm}^{-1} (c_{ijlm} + \delta_{il} \sigma_{jm}), \tag{43}$$

where $\boldsymbol{\sigma}$ is the Cauchy stress and $J = \det(\mathbf{F})$ is the Jacobian of deformation mapping. Then instability criteria corresponding to (37) is

$$\Lambda(\widetilde{\mathbf{u}}^0, \Xi) = (\widetilde{u}_I^0 \Xi_J) \widetilde{\mathbb{L}}_{IJLM} (\widetilde{u}_L^0 \Xi_M) \geq 0. \tag{44}$$

It is should be noticed that both major and minor symmetries, i.e., $(IJ \leftrightarrow LM)$ and $(I \leftrightarrow J)$, are posed by $C_{IJLM}$, while only the major symmetry is posed by $\widetilde{\mathbb{L}}_{IJLM}$. After analyses analogue to the condition (25), $\widetilde{\mathbb{L}}_{IJLM}$ could be expressed in a form with complete symmetries like $\mathbf{C}$ as

$$\widetilde{\mathbb{L}}_{IJLM}^S = \frac{1}{4} (\widetilde{\mathbb{L}}_{IJLM} + \widetilde{\mathbb{L}}_{JILM} + \widetilde{\mathbb{L}}_{IJML} + \widetilde{\mathbb{L}}_{JIML}).$$

(45)

Below, we consider the mechanical stabilities of crystals at current configuration. Assuming that the current configuration is under a small virtual disturbance of uniform deformation which brings current configuration into a new configuration, criteria (37) is still valid, which involve the current configuration and the new configuration instead. Namely, $\mathbf{C}$ and $\mathbf{S}$ in (41) correspond to elastic stiffness ($\mathbf{c}$) at current configuration and the Cauchy stress $\boldsymbol{\sigma}$. Because the virtual disturbance is small, we have

$$\mathbf{F} \approx \mathbf{I}, \ \mathbf{F}^{-1} \approx \mathbf{I}, \ J \approx 1. \tag{46}$$

Using results of (42) and (46), Eq. (45) becomes

$$\widetilde{\mathbb{L}}_{ijlm}^S = c_{ijlm} + \frac{1}{4} (\delta_{il} \sigma_{jm} + \delta_{jl} \sigma_{im} + \delta_{im} \sigma_{jl} + \delta_{jm} \sigma_{il}). \tag{47}$$

Instability condition (44) requires that $\widetilde{\mathbb{L}}^S$ should be positive semidefinite. In comparison with the modified Born criterion proposed by Wang et al[21,22], similar tensor is defined as

$$B_{IJLM}^S = c_{ijlm} + \frac{1}{2} (\delta_{il} \sigma_{jm} + \delta_{jl} \sigma_{im} + \delta_{im} \sigma_{jl} + \delta_{jm} \sigma_{il} - 2 \delta_{lm} \sigma_{ij}). \tag{48}$$

They are different by an additional term of $\frac{1}{4} (\delta_{il} \sigma_{jm} + \delta_{jl} \sigma_{im} + \delta_{im} \sigma_{jl} + \delta_{jm} \sigma_{il}) - \delta_{lm} \sigma_{ij}$. This is because $\mathbf{B}^S$ is derived by assuming that $\mathbf{F}$ is symmetric[22], while no this assumption is make in our derivations. For cubic crystals under homogenous pressures, the difference between (47) and (48) vanishes. However, in crystals with lower symmetries or under heterogeneous stress states, the modified Born criterion becomes incorrect due to the additional term.

Alternatively, considering Eq. (43) and (33), the instability criteria (44) could be rewritten as

$$\Lambda(\widetilde{\mathbf{u}}^0, \Xi) = (\widehat{u}_i^0 \widehat{\Xi}_j)(c_{ijlm} + \delta_{il} \sigma_{jm})(\widehat{u}_l^0 \widehat{\Xi}_m) = \widehat{\Xi}_j (c_{ijlm} \widehat{u}_i^0 \widehat{u}_l^0 + \|\widehat{\mathbf{u}}^0\|^2 \sigma_{jm}) \widehat{\Xi}_m \geq 0, \tag{49}$$

where $\widehat{\Xi} = \mathbf{F}^{-1}\Xi$ is the wavevector in current configuration. According to our previous analyses in Part II B, $\|\widehat{\mathbf{u}}^0\| = 1$. Then the above instability criterion reduce to the $\Lambda$-criterion proposed By Li et al [23]. From the derivations of (49), it can be concluded that mechanical instability criterion proposed by Li et al are equivalent with the one represented by (44), both of which are more precise than the modified Born criterion. According to (49), mechanical stabilities of crystals requires the minimum value $(\lambda_c^\Lambda)$ of $\Lambda$ to be non-negative for arbitrary $\widehat{\mathbf{u}}^0$ and $\widehat{\Xi}$. Since each tuple i.e., $(\widehat{\mathbf{u}}^0, \widehat{\Xi})$, corresponds to a planar wave mode due to atom collective vibrations, different critical values for $(\widehat{\mathbf{u}}^0, \widehat{\Xi})$, at which the minimum value of $\Lambda$ reached, indicate certain atom mechanisms when the mechanical instabilities take place. It is apparent that the instability criterion (49) does not rely on the magnitude of $\widehat{\Xi}$. Thus,

$$\lambda_c^\Lambda = MIN_{\|\widehat{\mathbf{u}}^0\|=1, \|\widehat{\Xi}\|=1} \Lambda(\widetilde{\mathbf{u}}^0, \Xi). \tag{50}$$

The above equation could be evaluated through a iterative procedure, designed by Li et al [23,24], as follows: Firstly, set initial a trial value of $\Xi$ to be (0, 0, 1) and calculate the minimum eigenvalue and the corresponding eigenvector of (50) at fixed $\Xi$; Secondly, replace $\widehat{\mathbf{u}}^0$ with the obtain eigenvector and minimize (50) at fixed $\widehat{\mathbf{u}}^0$. Then, use the eigenvector obtained in the last step as the value of $\Xi$ and repeat the first and second steps until convergence. However, this procedure cannot be applied for evaluating the GEIC because the latter is not a standard quadratic form.

## III. Microscopic comprehension of higher order elastic instabilities
### A. Microscopic expressions of elastic constants

Application of the GEIC to certain crystals requires to know the elastic constants defined by (17). Although the determination of strain-gradient elastic constants (SGEC), i.e., $^1\mathbb{H}$, $^1\mathbb{H}$ and $\mathbb{M}$, for crystals are nontrivial, significant progresses have been made to evaluate the SGECs for various crystals. For example, lattice dynamic approaches[8,13] are established to determine the SGE constants. Key point of the approach is to numerically fitting phonon dispersion relations along certain high symmetry directions in order to acquire the requisite elastic constants. However, great cautions should be taken when performing the fitting procedures. For instance, the fit should be carried out starting from k-vectors in the vicinity of zero to the one where dispersive effects just start to kick in [13]. Fitting at k-vector, corresponding to regions where frequencies are very high and dispersive effects are large, will results in spurious estimations of the elastic constants. Alternatively, a statistical mechanics approach, which relates the SGE constants to atomic displacement correlations in a molecular dynamics (MD) ensemble, is proposed by Maranganti et al[13]. This approach could be also applied for estimating the elastic constants of non-crystalline systems. To acquire the elastic constants with high precision, a large simulation cell should be employed in the MD simulations. Both approaches adopt a "dynamic" or "indirect" (statistical) way to acquire a proper estimation of the elastic constants. Most recently, a "direct" approach at atom level is proposed to obtain the SGECs by Admal et al[16]. Their derivations are based on condition of energetic equivalence between continuum and atomistic representations of crystals when kinetics of the latter is governed by the Cauchy-Born rule. Their starting point is based on a polynomial expansion of deformation map which is conceptually a continuum approach. In contrast, we determine the SGECs for crystals binding through arbitrary kinds of interatomic potentials using a direct method which is a generalization of the one proposed by Sunyk et al[19] for pair potentials. Results of ours are the same as the ones by Admal et al. Below, we use lowercase Greek letters, such as $\alpha, \beta, \gamma, \mu, \nu, \lambda$ and $\rho$, to distinguish the three Cartesian components of vectors or tensors, and lowercase English letters, such as $i, j, k$ ..., to stand for atom indexes. Summation over repeated indexes is only applied for the Cartesian indexes.

Extending the idea of Sunyk et al[19], generalized Cauchy-Born rule of third-order is assumed to be

$$\mathbf{r}^{\alpha\beta} = \mathbf{F} \cdot \mathbf{R}^{\alpha\beta} + \frac{1}{2}\mathbf{G} : (\mathbf{R}^{\alpha\beta} \otimes \mathbf{R}^{\alpha\beta}) + \frac{1}{6}\mathbf{H} \vdots (\mathbf{R}^{\alpha\beta} \otimes \mathbf{R}^{\alpha\beta} \otimes \mathbf{R}^{\alpha\beta}). \tag{51}$$

where $\mathbf{r}^{\alpha\beta} = \mathbf{r}^{\alpha} - \mathbf{r}^{\beta}$ and $\mathbf{R}^{\alpha\beta} = \mathbf{R}^{\alpha} - \mathbf{R}^{\beta}$ represent relative position vectors between atom $\alpha$ and $\beta$ in the deformed and initial configurations, respectively. In the next section, we will show that this assumption is equivalent to the higher order expansion of atom displacements over their neighbors. As mentioned in Part II A, under adiabatic conditions, strain energy density related to specific internal energy by

$$W(\mathbf{F}, \mathbf{G}) = \rho_0 U(\mathbf{r}^{\alpha}, s), \tag{52}$$

where $s$ is the specific entropy, $\rho_0 = 1/\Omega_{\mathcal{B}}$ and $\Omega_{\mathcal{B}}$ is the volume of a bulk of atoms of interest in the initial crystal. Since entropy does not change during elastic deformations, the specific internal energy could be replaced by potential energy of the atomic system when evaluate derivatives with respective to $\mathbf{F}$ and $\mathbf{G}$. Due to requirements of the invariance with respective to translations, rotations and reflections, the potential energy, i.e., $\mathcal{V}(\{r^{\alpha\beta}\})$, can be expressed as a function of relative distances between atoms for all interatomic potentials[25], for examples pair potentials (e.g. Morse and Lennard-Jones potentials), cluster potentials (e.g. three-body potentials), pair functionals (Embedded Atom Model (EAM) potentials and Finnis–Sinclair model potentials) and cluster functionals (e.g. Bond Order Potentials and Angular-dependent EAM potentials). According to definitions given in (17) and the generalized Cauchy-Born rule, we could obtain[**]

$$\mathbf{P} = \frac{1}{2\Omega_{\mathcal{B}}} \sum_{\substack{\alpha,\beta \\ \alpha \neq \beta}} \varphi_{\alpha\beta} \frac{\mathbf{r}^{\alpha\beta}}{r^{\alpha\beta}} \otimes \mathbf{R}^{\alpha\beta}, \tag{53}$$

$$\mathbb{L} = \frac{1}{2\Omega_{\mathcal{B}}} \left\{ \frac{1}{2} \sum_{\substack{\alpha,\beta,\mu,\nu \\ \alpha \neq \beta \\ \mu \neq \nu}} \kappa_{\alpha\beta\mu\nu} \frac{\mathbf{r}^{\alpha\beta}}{r^{\alpha\beta}} \otimes \mathbf{R}^{\alpha\beta} \otimes \frac{\mathbf{r}^{\mu\nu}}{r^{\mu\nu}} \otimes \mathbf{R}^{\mu\nu} \right.$$

$$\left. + \sum_{\substack{\alpha,\beta \\ \alpha \neq \beta}} \varphi_{\alpha\beta} \frac{1}{r^{\alpha\beta}} \left( \mathbf{I} - \frac{\mathbf{r}^{\alpha\beta} \otimes \mathbf{r}^{\alpha\beta}}{r^{\alpha\beta} r^{\alpha\beta}} \right) \overline{\otimes} (\mathbf{R}^{\alpha\beta} \otimes \mathbf{R}^{\alpha\beta}) \right\},$$

$$\tag{54}$$

where $\varphi_{\alpha\beta}$ and $\kappa_{\alpha\beta\mu\nu}$ are bond force and bond stiffness, which are defined by

$$\varphi_{\alpha\beta} \equiv \frac{\partial \mathcal{V}}{\partial r^{\alpha\beta}} \text{ and } \kappa_{\alpha\beta\mu\nu} = \frac{\partial^2 \mathcal{V}}{\partial r^{\alpha\beta} \partial r^{\mu\nu}}, \tag{55}$$

respectively. The derivations above have employed the differential identities below:

$$\frac{\partial r^{\alpha\beta}}{\partial r^{\mu\nu}} = \delta_{\alpha\mu}\delta_{\beta\nu} + \delta_{\alpha\nu}\delta_{\beta\mu}, \tag{56}$$

---

** For arbitrary rank-two tensor $\mathbf{A}$ and $\mathbf{B}$, the non-standard dyadic product $\overline{\otimes}$ and $\underline{\otimes}$ between them are $[\mathbf{A} \overline{\otimes} \mathbf{B}]_{ijkl} \doteq A_{ik}B_{jl}$, $[\mathbf{A} \underline{\otimes} \mathbf{B}]_{ijkl} \doteq A_{il}B_{jk}$. And for arbitary rank-three tensor $\mathbf{Q}$, the non-standard dyadic product $\underline{\underline{\otimes}}$ is defined by $[\mathbf{A} \underline{\underline{\otimes}} \mathbf{Q}]_{ijklm} \doteq A_{im}Q_{jkl}$.

$$\frac{\partial r^{\alpha\beta}}{\partial \mathbf{r}^{\alpha\beta}} = \frac{\mathbf{r}^{\alpha\beta}}{r^{\alpha\beta}}. \tag{57}$$

Similarly, using Eq. (17) and the generalized Cauchy-Born rule, the microscopic expressions for the SGECs is obtained as follows:

$$\mathbf{Q} = \frac{1}{4\Omega_{\mathcal{B}}} \sum_{\substack{\alpha,\beta \\ \alpha \neq \beta}} \varphi_{\alpha\beta} \frac{\mathbf{r}^{\alpha\beta}}{r^{\alpha\beta}} \otimes \mathbf{R}^{\alpha\beta} \otimes \mathbf{R}^{\alpha\beta}, \tag{58}$$

$$\mathbb{H} = \frac{1}{4\Omega_{\mathcal{B}}} \Bigg\{ \frac{1}{2} \sum_{\substack{\alpha,\beta,\mu,\nu \\ \alpha \neq \beta \\ \mu \neq \nu}} \kappa_{\alpha\beta\mu\nu} \frac{\mathbf{r}^{\alpha\beta}}{r^{\alpha\beta}} \otimes \mathbf{R}^{\alpha\beta} \otimes \frac{\mathbf{r}^{\mu\nu}}{r^{\mu\nu}} \otimes \mathbf{R}^{\mu\nu} \otimes \mathbf{R}^{\mu\nu}$$

$$+ \sum_{\substack{\alpha,\beta \\ \alpha \neq \beta}} \varphi_{\alpha\beta} \frac{1}{r^{\alpha\beta}} \left( \mathbf{I} - \frac{\mathbf{r}^{\alpha\beta} \otimes \mathbf{r}^{\alpha\beta}}{r^{\alpha\beta} r^{\alpha\beta}} \right) \overline{\otimes} \left( \mathbf{R}^{\alpha\beta} \otimes \mathbf{R}^{\alpha\beta} \right) \otimes \mathbf{R}^{\alpha\beta} \Bigg\}, \tag{59}$$

$$\mathbb{H}^{\ddagger} = \frac{1}{4\Omega_{\mathcal{B}}} \Bigg\{ \frac{1}{2} \sum_{\substack{\alpha,\beta,\mu,\nu \\ \alpha \neq \beta \\ \mu \neq \nu}} \kappa_{\alpha\beta\mu\nu} \frac{\mathbf{r}^{\alpha\beta}}{r^{\alpha\beta}} \otimes \mathbf{R}^{\alpha\beta} \otimes \mathbf{R}^{\alpha\beta} \otimes \frac{\mathbf{r}^{\mu\nu}}{r^{\mu\nu}} \otimes \mathbf{R}^{\mu\nu}$$

$$+ \sum_{\substack{\alpha,\beta \\ \alpha \neq \beta}} \varphi_{\alpha\beta} \frac{1}{r^{\alpha\beta}} \left( \mathbf{I} - \frac{\mathbf{r}^{\alpha\beta} \otimes \mathbf{r}^{\alpha\beta}}{r^{\alpha\beta} r^{\alpha\beta}} \right) \underline{\otimes} \left( \mathbf{R}^{\alpha\beta} \otimes \mathbf{R}^{\alpha\beta} \right) \otimes \mathbf{R}^{\alpha\beta} \Bigg\}, \tag{60}$$

$$\mathbb{M} = \frac{1}{8\Omega_{\mathcal{B}}} \Bigg\{ \frac{1}{2} \sum_{\substack{\alpha,\beta,\mu,\nu \\ \alpha \neq \beta \\ \mu \neq \nu}} \kappa_{\alpha\beta\mu\nu} \frac{\mathbf{r}^{\alpha\beta}}{r^{\alpha\beta}} \otimes \mathbf{R}^{\alpha\beta} \otimes \mathbf{R}^{\alpha\beta} \otimes \frac{\mathbf{r}^{\mu\nu}}{r^{\mu\nu}} \otimes \mathbf{R}^{\mu\nu} \otimes \mathbf{R}^{\mu\nu}$$

$$+ \sum_{\substack{\alpha,\beta \\ \alpha \neq \beta}} \varphi_{\alpha\beta} \frac{1}{r^{\alpha\beta}} \left( \mathbf{I} - \frac{\mathbf{r}^{\alpha\beta} \otimes \mathbf{r}^{\alpha\beta}}{r^{\alpha\beta} r^{\alpha\beta}} \right) \underline{\otimes} \left( \mathbf{R}^{\alpha\beta} \otimes \mathbf{R}^{\alpha\beta} \right) \otimes \left( \mathbf{R}^{\alpha\beta} \otimes \mathbf{R}^{\alpha\beta} \right) \Bigg\}, \tag{61}$$

$$\mathbb{E} = \frac{1}{12\Omega_{\mathcal{B}}} \Bigg\{ \frac{1}{2} \sum_{\substack{\alpha,\beta,\mu,\nu \\ \alpha \neq \beta \\ \mu \neq \nu}} \kappa_{\alpha\beta\mu\nu} \frac{\mathbf{r}^{\alpha\beta}}{r^{\alpha\beta}} \otimes \mathbf{R}^{\alpha\beta} \otimes \frac{\mathbf{r}^{\mu\nu}}{r^{\mu\nu}} \otimes \mathbf{R}^{\mu\nu} \otimes \mathbf{R}^{\mu\nu} \otimes \mathbf{R}^{\mu\nu}$$

$$+ \sum_{\substack{\alpha,\beta \\ \alpha \neq \beta}} \varphi_{\alpha\beta} \frac{1}{r^{\alpha\beta}} \left( \mathbf{I} - \frac{\mathbf{r}^{\alpha\beta} \otimes \mathbf{r}^{\alpha\beta}}{r^{\alpha\beta} r^{\alpha\beta}} \right) \overline{\otimes} \left( \mathbf{R}^{\alpha\beta} \otimes \mathbf{R}^{\alpha\beta} \right) \otimes \mathbf{R}^{\alpha\beta} \otimes \mathbf{R}^{\alpha\beta} \Bigg\}, \tag{62}$$

Because $\kappa_{\alpha\beta\mu\nu}$ holds symmetries of $(\alpha \leftrightarrow \beta)$ and $(\mu \leftrightarrow \nu)$, it is easy to show that the first summation term in Eq. (58-61) vanishes by noting relations of $\mathbf{R}^{\alpha\beta} = -\mathbf{R}^{\beta\alpha}$. Specially, $\mathbb{M}$ is in fact only $3/2$ times of the second summation term of $\mathbb{E}$ and thus, $\mathbb{E}$ would give rise to a major correction to the higher order stabilities (See Eq. 32) through $\mathbb{W}$. From Eq. (34), we have

$$\tilde{\mathbb{L}} = \frac{1}{2\Omega_{\mathcal{B}}} \left\{ \frac{1}{2} \sum_{\substack{\alpha,\beta,\mu,\nu \\ \alpha \neq \beta \\ \mu \neq \nu}} \kappa_{\alpha\beta\mu\nu} \frac{\overline{\mathbf{R}}^{\alpha\beta}}{\overline{R}^{\alpha\beta}} \otimes \mathbf{R}^{\alpha\beta} \otimes \frac{\overline{\mathbf{R}}^{\mu\nu}}{\overline{R}^{\mu\nu}} \otimes \mathbf{R}^{\mu\nu} \right.$$

$$\left. + \sum_{\substack{\alpha,\beta \\ \alpha \neq \beta}} \varphi_{\alpha\beta} \frac{1}{\overline{R}^{\alpha\beta}} \left( \mathbf{I} - \frac{\overline{\mathbf{R}}^{\alpha\beta} \otimes \overline{\mathbf{R}}^{\alpha\beta}}{\overline{R}^{\alpha\beta}\overline{R}^{\alpha\beta}} \right) \overline{\otimes} \left( \mathbf{R}^{\alpha\beta} \otimes \mathbf{R}^{\alpha\beta} \right) \right\},$$

(63)

$$\tilde{\mathbb{M}} = \frac{1}{8\Omega_{\mathcal{B}}} \left\{ \frac{1}{2} \sum_{\substack{\alpha,\beta,\mu,\nu \\ \alpha \neq \beta \\ \mu \neq \nu}} \kappa_{\alpha\beta\mu\nu} \frac{\overline{\mathbf{R}}^{\alpha\beta}}{\overline{R}^{\alpha\beta}} \otimes \mathbf{R}^{\alpha\beta} \otimes \mathbf{R}^{\alpha\beta} \otimes \frac{\overline{\mathbf{R}}^{\mu\nu}}{\overline{R}^{\mu\nu}} \otimes \mathbf{R}^{\mu\nu} \otimes \mathbf{R}^{\mu\nu} \right.$$

$$\left. + \sum_{\substack{\alpha,\beta \\ \alpha \neq \beta}} \varphi_{\alpha\beta} \frac{1}{\overline{R}^{\alpha\beta}} \left( \mathbf{I} - \frac{\overline{\mathbf{R}}^{\alpha\beta} \otimes \overline{\mathbf{R}}^{\alpha\beta}}{\overline{R}^{\alpha\beta}\overline{R}^{\alpha\beta}} \right) \underline{\otimes} \left( \mathbf{R}^{\alpha\beta} \otimes \mathbf{R}^{\alpha\beta} \right) \otimes \left( \mathbf{R}^{\alpha\beta} \otimes \mathbf{R}^{\alpha\beta} \right) \right\},$$

(64)

where $\overline{\mathbf{R}}^{\alpha\beta} = \mathbf{F}^{-1}\mathbf{r}^{\alpha\beta}$. It should be noted that $\overline{\mathbf{R}}^{\alpha\beta}$ does not equal to $\mathbf{R}^{\alpha\beta}$ because of the presence of the higher-order deformation gradients. However, when the higher-order deformation gradients are small, the tinny difference between them could be neglected. This is the very case frequently encountered in the mechanical stability analyses of crystals, where initial crystals are assumed to be deformed by a small virtual disturbance of $\mathbf{F}$ and $\mathbf{G}$. In this case, Eq. (46) is valid so that the relative position vectors ($\mathbf{R}^{\alpha\beta}$) in Eq. (53), (54) and (58-62) could be replaced by $\mathbf{r}^{\alpha\beta}$. Thus, all elastic constants in the current configuration could be determined when interatomic potentials for crystals of interests are known.

In the next part of present work, the higher-order instability criterion will be employed to investigate the mechanical instabilities of several metals binding through EAM potentials. According to framework of embedded atom models, total potential energy of an atomic system could be expressed by the sum of total interatomic interactions and embedding energy, that is,

$$\mathcal{V} = \frac{1}{2} \sum_{\substack{\alpha,\beta \\ \alpha \neq \beta}} \phi(r^{\alpha\beta}) + \sum_{\alpha} F(\rho_\alpha), \tag{65}$$

where total electron density $\rho_\alpha$ at $\mathbf{r}^\alpha$ is contributed by spherically averaged atom electron density $f(r^{\alpha\beta})$ from surrounding atoms, that is

$$\rho_\alpha = \sum_{\substack{\beta \\ \beta \neq \alpha}} f(r^{\alpha\beta}). \tag{66}$$

The summations in Eq. (65) and (66) run over all neighbors within a distance of $r_c$ from the central atom. Hereafter, we refer to the distance as cutoff distance. Detailed function form of pairwise interaction $\phi(r)$, atom electron density $f(r)$ and embedding energy $F(\rho)$ could be given in either tabulated or analytic forms. From Eq. (55) and (65), the bond force and bond stiffness for the EAM potentials are

$$\varphi_{\alpha\beta} = \phi'(r^{\alpha\beta}) + [F'(\rho_\alpha) + F'(\rho_\beta)]f'(r^{\alpha\beta}), \tag{67}$$

$$\kappa_{\alpha\beta\mu\nu} = \phi''(r^{\alpha\beta})(\delta_{\alpha\mu}\delta_{\beta\nu} + \delta_{\alpha\nu}\delta_{\beta\mu}) + f'(r^{\alpha\beta})F''(\rho_\alpha)[f'(r^{\alpha\nu})\delta_{\alpha\mu} + f'(r^{\alpha\mu})\delta_{\alpha\nu}]$$
$$+ f'(r^{\alpha\beta})F''(\rho_\beta)[f'(r^{\beta\nu})\delta_{\beta\mu} + f'(r^{\beta\mu})\delta_{\beta\nu}]$$
$$+ [F'(\rho_\alpha) + F'(\rho_\beta)]f''(r^{\alpha\beta})(\delta_{\alpha\mu}\delta_{\beta\nu} + \delta_{\alpha\nu}\delta_{\beta\mu}).$$

(68)

Substituting the above results into Eq. (63) and (64), detailed expressions for $\widetilde{\mathbb{L}}$ and $\widetilde{\mathbb{M}}$ could be obtained. Simplified microscopic expressions for $\widetilde{\mathbb{L}}$ and $\widetilde{\mathbb{M}}$ in their component forms could be found in Appendix A. The other elastic constants could be calculated in similar manners. Using above microscopic expressions, the independent components of $\widetilde{\mathbb{M}}$ are calculated and listed in Table I, which agrees well with results evaluated through method proposed by Admal et al[16]. And three additional generalized Cauchy relationships can be found in our results, while they only approximately satisfied in the results of Admal et al. More details could be found in Appendix A.

**B. Microscopic comprehension on the higher-order phenomenological theory**

Imaging that a volume ($\Omega_{\mathfrak{B}}$) in a crystal, consisting of $\mathcal{N}$ atoms, is deformed from reference configuration $\{\mathbf{R}^\gamma\}$ to current configuration $\{\mathbf{r}^\gamma\}$ under a small virtual deformation, we will evaluate all elastic constants for the volume. Note that the deformation does not necessarily to be uniform over the volume. Relative position vector between atom $\alpha$ and $\beta$ in configuration $\{\mathbf{R}^\gamma\}$ and $\{\mathbf{r}^\gamma\}$ are denoted by $\mathbf{R}^{\alpha\beta} = \mathbf{R}^\alpha - \mathbf{R}^\beta$ and $\mathbf{r}^{\alpha\beta} = \mathbf{r}^\alpha - \mathbf{r}^\beta$, respectively. Displacement of atom $\alpha$ at $\mathbf{R}^\alpha$ could be expanded at the position of its neighbors (For example atom $\alpha$), to the fourth order, in terms of the relative position vector between them, that is

$$\mathbf{u}^\alpha = \mathbf{u}^\beta + (\mathbf{u}^\beta \nabla)\mathbf{R}^{\alpha\beta} + \frac{1}{2}(\mathbf{u}^\beta \nabla \nabla):(\mathbf{R}^{\alpha\beta} \otimes \mathbf{R}^{\alpha\beta}) + \frac{1}{6}(\mathbf{u}^\beta \nabla \nabla \nabla)$$
$$\vdots (\mathbf{R}^{\alpha\beta} \otimes \mathbf{R}^{\alpha\beta} \otimes \mathbf{R}^{\alpha\beta}) + \frac{1}{24}(\mathbf{u}^\beta \nabla \nabla \nabla \nabla) \vdots (\mathbf{R}^{\alpha\beta} \otimes \mathbf{R}^{\alpha\beta} \otimes \mathbf{R}^{\alpha\beta} \otimes \mathbf{R}^{\alpha\beta}),$$

(69)

where $\mathbf{u}^\alpha = \mathbf{u}(\mathbf{R}^\alpha) = \mathbf{r}^\alpha - \mathbf{R}^\alpha$. This expansion lays the physical bases of the generalized Cauchy-Born rule (See Eq. 51), which is precise enough for metals since effective interactions between atoms are short-ranged due to Coulomb screening effects. Recently, it has been justified[6] in single crystalline iron under ramp compressions up to a strain rate of $10^{12}$ s$^{-1}$. Supposing that displacements of atoms are small compared with lattice constant, which is generally true for stable crystals below melting points, specific potential energy of the deformed crystal can be expanded about $\{\mathbf{R}^\gamma\}$, to the second order:

$$U(\{\mathbf{r}^\gamma\}) = U_0(\{\mathbf{R}^\gamma\}) - \sum_{\alpha=1}^{\mathcal{N}} \mathbf{f}^\alpha \mathbf{u}^\alpha + \frac{1}{2}\sum_{\alpha=1}^{\mathcal{N}}\sum_{\beta=1}^{\mathcal{N}} \mathbf{\Phi}^{\alpha\beta}:(\mathbf{u}^\alpha \otimes \mathbf{u}^\beta),$$ (70)

where force on atom $\alpha$ and force constant are defined by

$$\mathbf{f}^\alpha = -\left.\frac{\partial U}{\partial \mathbf{u}^\alpha}\right|_{\{\mathbf{R}^\gamma\}} \quad \text{and} \quad \mathbf{\Phi}^{\alpha\beta} = \left.\frac{\partial^2 U}{\partial \mathbf{u}^\alpha \partial \mathbf{u}^\beta}\right|_{\{\mathbf{R}^\gamma\}}.$$ (71)

Eq. (70) could be further expressed in terms of summations of over $\alpha - \beta$ pairs by using the acoustic sum rule, i.e., $\Phi_{ij}^{\alpha\alpha} = -\sum_{\substack{\beta \neq \alpha \\ \beta \in \mathfrak{B}}} \Phi_{ij}^{\alpha\beta}$, that is

$$U(\{\mathbf{r}^\gamma\}) = U_0(\{\mathbf{R}^\gamma\}) - \sum_\alpha \mathbf{f}^\alpha \mathbf{u}^\alpha + \frac{1}{2}\sum_\alpha \sum_{\beta \neq \alpha} \mathbf{\Phi}^{\alpha\beta}:\left(\mathbf{u}^\alpha \otimes (\mathbf{u}^\beta - \mathbf{u}^\alpha)\right),$$ (72)

or in a more symmetric form of

$$U(\{\mathbf{r}^\gamma\}) = U_0(\{\mathbf{R}^\gamma\}) + \frac{1}{2}\sum_\alpha \sum_{\beta\neq\alpha} \boldsymbol{\varphi}^{\alpha\beta}(\mathbf{u}^\beta - \mathbf{u}^\alpha)$$
$$- \frac{1}{4}\sum_\alpha \sum_{\beta\neq\alpha} \boldsymbol{\Phi}^{\alpha\beta}:\left((\mathbf{u}^\beta - \mathbf{u}^\alpha)\otimes(\mathbf{u}^\beta - \mathbf{u}^\alpha)\right),$$
(73)

where $\boldsymbol{\Phi}^{\alpha\beta} = \boldsymbol{\Phi}^{\beta\alpha}$ and $\mathbf{f}^\alpha = \sum_{\beta\neq\alpha} \boldsymbol{\varphi}^{\alpha\beta}$ is employed. And $\boldsymbol{\varphi}^{\alpha\beta}$ is the force on atom $\alpha$ due to atom $\beta$, which is defined by

$$\boldsymbol{\varphi}^{\alpha\beta} = \left.\frac{\partial U}{\partial \mathbf{r}^{\alpha\beta}}\right|_{\{\mathbf{R}^\gamma\}}. \tag{74}$$

Through substituting Eq. (69) into Eq. (73) and rearranging the resulting expression according to orders of the first and second derivatives of displacements, the specific energy due to the small deformation disturbance is

$$U(\{\mathbf{r}^\gamma\}) = U_0(\{\mathbf{R}^\gamma\}) + \frac{1}{2}\sum_\alpha \left[\sum_{\beta\neq\alpha}\left(\boldsymbol{\varphi}^{\alpha\beta} \otimes \mathbf{R}^{\beta\alpha}\right)\right]:(\mathbf{u}^\alpha \nabla)$$

$$+ \frac{1}{4}\sum_\alpha \left[\sum_{\beta\neq\alpha}\left(\boldsymbol{\varphi}^{\alpha\beta} \otimes \mathbf{R}^{\beta\alpha} \otimes \mathbf{R}^{\beta\alpha}\right)\right] \vdots (\mathbf{u}^\alpha \nabla\nabla)$$

$$+ \frac{1}{12}\sum_\alpha \left[\sum_{\beta\neq\alpha}\left(\boldsymbol{\varphi}^{\alpha\beta} \otimes \mathbf{R}^{\beta\alpha} \otimes \mathbf{R}^{\beta\alpha} \otimes \mathbf{R}^{\beta\alpha}\right)\right] \vdots\!\vdots (\mathbf{u}^\alpha \nabla\nabla\nabla)$$

$$+ \frac{1}{48}\sum_\alpha \left[\sum_{\beta\neq\alpha}\left(\boldsymbol{\varphi}^{\alpha\beta} \otimes \mathbf{R}^{\beta\alpha} \otimes \mathbf{R}^{\beta\alpha} \otimes \mathbf{R}^{\beta\alpha} \otimes \mathbf{R}^{\beta\alpha}\right)\right] ::: (\mathbf{u}^\alpha \nabla\nabla\nabla\nabla)$$

$$- \frac{1}{4}\sum_\alpha (\mathbf{u}^\alpha \nabla):\left[\sum_{\beta\neq\alpha}\left(\boldsymbol{\Phi}^{\alpha\beta} \overline{\otimes} \left(\mathbf{R}^{\beta\alpha} \otimes \mathbf{R}^{\beta\alpha}\right)\right)\right]:(\mathbf{u}^\alpha \nabla)$$

$$- \frac{1}{8}\sum_\alpha (\mathbf{u}^\alpha \nabla):\left[\sum_{\beta\neq\alpha}\left(\boldsymbol{\Phi}^{\alpha\beta} \overline{\otimes} \left(\mathbf{R}^{\beta\alpha} \otimes \mathbf{R}^{\beta\alpha}\right) \otimes \mathbf{R}^{\beta\alpha}\right)\right] \vdots (\mathbf{u}^\alpha \nabla\nabla)$$

$$- \frac{1}{8}\sum_\alpha (\mathbf{u}^\alpha \nabla\nabla) \vdots \left[\sum_{\beta\neq\alpha}\left(\boldsymbol{\Phi}^{\alpha\beta} \underline{\otimes} \left(\mathbf{R}^{\beta\alpha} \otimes \mathbf{R}^{\beta\alpha}\right) \otimes \mathbf{R}^{\beta\alpha}\right)\right]:(\mathbf{u}^\alpha \nabla)$$

$$- \frac{1}{16}\sum_\alpha (\mathbf{u}^\alpha \nabla\nabla) \vdots \left[\sum_{\beta\neq\alpha}\left(\boldsymbol{\Phi}^{\alpha\beta} \underline{\otimes} \left(\mathbf{R}^{\beta\alpha} \otimes \mathbf{R}^{\beta\alpha}\right) \otimes \mathbf{R}^{\beta\alpha} \otimes \mathbf{R}^{\beta\alpha}\right)\right] \vdots (\mathbf{u}^\alpha \nabla\nabla)$$

$$- \frac{1}{24}\sum_\alpha (\mathbf{u}^\alpha \nabla):\left[\sum_{\beta\neq\alpha}\left(\boldsymbol{\Phi}^{\alpha\beta} \underline{\otimes} \left(\mathbf{R}^{\beta\alpha} \otimes \mathbf{R}^{\beta\alpha}\right) \otimes \mathbf{R}^{\beta\alpha} \otimes \mathbf{R}^{\beta\alpha}\right)\right] \vdots\!\vdots (\mathbf{u}^\alpha \nabla\nabla\nabla)$$

$$- \frac{1}{24}\sum_\alpha (\mathbf{u}^\alpha \nabla\nabla\nabla) \vdots\!\vdots \left[\sum_{\beta\neq\alpha}\left(\boldsymbol{\Phi}^{\alpha\beta} \underline{\underline{\otimes}} \left(\mathbf{R}^{\beta\alpha} \otimes \mathbf{R}^{\beta\alpha} \otimes \mathbf{R}^{\beta\alpha}\right) \otimes \mathbf{R}^{\beta\alpha}\right)\right]:(\mathbf{u}^\alpha \nabla).$$
(75)

To connect the specific energy with the strain energy density, average specific energy density ($\mathcal{U}^\alpha$)

over a characteristic volume $\Omega_{\mathcal{B}}$ centered at atom $\alpha$ is $\mathcal{U}^\alpha = \overline{U}/\Omega_{\mathcal{B}}$ where $\overline{U}$ is assumed to be expressed by

$$\overline{U}(\{\mathbf{r}^\gamma\}) = \overline{U}_0(\{\mathbf{R}^\gamma\}) + \frac{1}{2}\left[\sum_\alpha \sum_{\beta \neq \alpha} \left(\boldsymbol{\varphi}^{\alpha\beta} \otimes \mathbf{R}^{\beta\alpha}\right)\right] : (\overline{\mathbf{u}}\nabla)$$

$$+ \frac{1}{4}\left[\sum_\alpha \sum_{\beta \neq \alpha} \left(\boldsymbol{\varphi}^{\alpha\beta} \otimes \mathbf{R}^{\beta\alpha} \otimes \mathbf{R}^{\beta\alpha}\right)\right] : (\overline{\mathbf{u}}\nabla\nabla)$$

$$+ \frac{1}{12}\left[\sum_\alpha \sum_{\beta \neq \alpha} \left(\boldsymbol{\varphi}^{\alpha\beta} \otimes \mathbf{R}^{\beta\alpha} \otimes \mathbf{R}^{\beta\alpha} \otimes \mathbf{R}^{\beta\alpha}\right)\right] \vdots (\overline{\mathbf{u}}\nabla\nabla\nabla)$$

$$+ \frac{1}{48}\left[\sum_\alpha \sum_{\beta \neq \alpha} \left(\boldsymbol{\varphi}^{\alpha\beta} \otimes \mathbf{R}^{\beta\alpha} \otimes \mathbf{R}^{\beta\alpha} \otimes \mathbf{R}^{\beta\alpha} \otimes \mathbf{R}^{\beta\alpha}\right)\right] :: (\overline{\mathbf{u}}\nabla\nabla\nabla\nabla)$$

$$- \frac{1}{4}(\overline{\mathbf{u}}\nabla) : \left[\sum_\alpha \sum_{\beta \neq \alpha} \left(\boldsymbol{\Phi}^{\alpha\beta} \overline{\otimes} \left(\mathbf{R}^{\beta\alpha} \otimes \mathbf{R}^{\beta\alpha}\right)\right)\right] : (\overline{\mathbf{u}}\nabla)$$

$$- \frac{1}{8}(\overline{\mathbf{u}}\nabla) : \left[\sum_\alpha \sum_{\beta \neq \alpha} \left(\boldsymbol{\Phi}^{\alpha\beta} \overline{\otimes} \left(\mathbf{R}^{\beta\alpha} \otimes \mathbf{R}^{\beta\alpha}\right) \otimes \mathbf{R}^{\beta\alpha}\right)\right] \vdots (\overline{\mathbf{u}}\nabla\nabla)$$

$$- \frac{1}{8}(\overline{\mathbf{u}}\nabla\nabla) : \left[\sum_\alpha \sum_{\beta \neq \alpha} \left(\boldsymbol{\Phi}^{\alpha\beta} \underline{\otimes} \left(\mathbf{R}^{\beta\alpha} \otimes \mathbf{R}^{\beta\alpha}\right) \otimes \mathbf{R}^{\beta\alpha}\right)\right] : (\overline{\mathbf{u}}\nabla)$$

$$- \frac{1}{16}(\overline{\mathbf{u}}\nabla\nabla) : \left[\sum_\alpha \sum_{\beta \neq \alpha} \left(\boldsymbol{\Phi}^{\alpha\beta} \underline{\otimes} \left(\mathbf{R}^{\beta\alpha} \otimes \mathbf{R}^{\beta\alpha}\right) \otimes \mathbf{R}^{\beta\alpha} \otimes \mathbf{R}^{\beta\alpha}\right)\right] : (\overline{\mathbf{u}}\nabla\nabla)$$

$$- \frac{1}{24}(\overline{\mathbf{u}}\nabla) : \left[\sum_\alpha \sum_{\beta \neq \alpha} \left(\boldsymbol{\Phi}^{\alpha\beta} \overline{\otimes} \left(\mathbf{R}^{\beta\alpha} \otimes \mathbf{R}^{\beta\alpha}\right) \otimes \mathbf{R}^{\beta\alpha} \otimes \mathbf{R}^{\beta\alpha}\right)\right] \vdots (\overline{\mathbf{u}}\nabla\nabla\nabla)$$

$$- \frac{1}{24}(\overline{\mathbf{u}}\nabla\nabla\nabla) \vdots \left[\sum_\alpha \sum_{\beta \neq \alpha} \left(\boldsymbol{\Phi}^{\alpha\beta} \underline{\underline{\otimes}} \left(\mathbf{R}^{\beta\alpha} \otimes \mathbf{R}^{\beta\alpha} \otimes \mathbf{R}^{\beta\alpha}\right) \otimes \mathbf{R}^{\beta\alpha}\right)\right] : (\overline{\mathbf{u}}\nabla).$$

(76)

The above expression could be interpreted by the below two cases: i) If the crystal is disturbed by a uniform strain and the characteristic volume is arbitrary large, the above equation reduces to Eq. (75) by noting that $\mathbf{u}^\alpha\nabla = \overline{\mathbf{u}}\nabla$, $\mathbf{u}^\alpha\nabla\nabla = \overline{\mathbf{u}}\nabla\nabla = \mathbf{0}$, $\mathbf{u}^\alpha\nabla\nabla\nabla = \overline{\mathbf{u}}\nabla\nabla\nabla = \mathbf{0}$ and $\mathbf{u}^\alpha\nabla\nabla\nabla\nabla = \overline{\mathbf{u}}\nabla\nabla\nabla\nabla = \mathbf{0}$; ii) if the crystal is disturbed by a uniform strain gradient and the characteristic volume is taken to be average atom volume, the specific energies at $\mathbf{r}^\gamma$ evaluated by the two expressions are the same by noting that $\mathbf{u}^\alpha\nabla = \overline{\mathbf{u}}\nabla$, $\mathbf{u}^\alpha\nabla\nabla = \overline{\mathbf{u}}\nabla\nabla$, $\mathbf{u}^\alpha\nabla\nabla\nabla = \overline{\mathbf{u}}\nabla\nabla\nabla = \mathbf{0}$ and $\mathbf{u}^\alpha\nabla\nabla\nabla\nabla = \overline{\mathbf{u}}\nabla\nabla\nabla\nabla = \mathbf{0}$. In general, $\overline{\mathbf{u}}\nabla$, $\overline{\mathbf{u}}\nabla\nabla$, $\overline{\mathbf{u}}\nabla\nabla\nabla$ and $\overline{\mathbf{u}}\nabla\nabla\nabla\nabla$ represents average of $\mathbf{u}^\alpha\nabla$, $\mathbf{u}^\alpha\nabla\nabla$, $\mathbf{u}^\alpha\nabla\nabla\nabla$ and $\mathbf{u}^\alpha\nabla\nabla\nabla\nabla$ over the characteristic volume which eliminates the difference between energy evaluated by Eq. (75) and (76). At continuum level, average strain energy density over $\Omega_{\mathcal{B}}$ could be expressed into homogeneous differential expansions with respective to the displacements, up to the fourth order, that is,

$$W(\mathbf{F},\mathbf{G}) = \mathbf{P}{:}\mathbf{F} + \mathbf{Q} \vdots \mathbf{G} + \mathbf{T} \vdots\!\vdots \mathbf{H} + \mathbf{\Gamma}{::}\mathbf{\Pi} + \frac{1}{2}\mathbf{F}{:}\mathbb{L}{:}\mathbf{F} + \mathbf{F}{:}\,^1\mathbb{H} \vdots \mathbf{G} + \mathbf{G} \vdots\,^2\mathbb{H}{:}\mathbf{F} + \frac{1}{2}\mathbf{G} \vdots \mathbb{M}$$

$$\vdots \mathbf{G}\ + \mathbf{F}{:}\,^1\mathbb{E} \vdots\!\vdots \mathbf{H} + \mathbf{H} \vdots\!\vdots\,^2\mathbb{E}{:}\mathbf{F}. \tag{77}$$

where $\mathbf{T} = \partial W/\partial \mathbf{H}$, $\mathbf{\Pi} = \mathbf{u}\nabla\nabla\nabla$ and $\mathbf{\Gamma} = \partial W/\partial \mathbf{\Pi}$. By identifying the expression of $\mathcal{U}^\alpha$ with the above one for $W$ and using the definitions below

$$\mathbf{F} = (\bar{\mathbf{u}}\nabla),\ \mathbf{G} = \bar{\mathbf{u}}\nabla\nabla,\ \mathbf{H} = \bar{\mathbf{u}}\nabla\nabla\nabla\ \text{and}\ \mathbf{\Pi} = \bar{\mathbf{u}}\nabla\nabla\nabla\nabla, \tag{78}$$

we obtain expressions of the elastic constants below:

$$\mathbf{P} = \frac{1}{2\Omega_\mathfrak{B}} \sum_\alpha \sum_{\beta\neq\alpha} \big(\boldsymbol{\varphi}^{\alpha\beta} \otimes \mathbf{R}^{\beta\alpha}\big), \tag{79}$$

$$\mathbf{Q} = \frac{1}{4\Omega_\mathfrak{B}} \sum_\alpha \sum_{\beta\neq\alpha} \big(\boldsymbol{\varphi}^{\alpha\beta} \otimes \mathbf{R}^{\beta\alpha} \otimes \mathbf{R}^{\beta\alpha}\big), \tag{80}$$

$$\mathbf{T} = \frac{1}{12\Omega_\mathfrak{B}} \Big[\sum_\alpha \sum_{\beta\neq\alpha} \big(\boldsymbol{\varphi}^{\alpha\beta} \otimes \mathbf{R}^{\beta\alpha} \otimes \mathbf{R}^{\beta\alpha} \otimes \mathbf{R}^{\beta\alpha}\big)\Big], \tag{81}$$

$$\mathbf{\Gamma} = \frac{1}{48\Omega_\mathfrak{B}} \Big[\sum_\alpha \sum_{\beta\neq\alpha} \big(\boldsymbol{\varphi}^{\alpha\beta} \otimes \mathbf{R}^{\beta\alpha} \otimes \mathbf{R}^{\beta\alpha} \otimes \mathbf{R}^{\beta\alpha} \otimes \mathbf{R}^{\beta\alpha}\big)\Big], \tag{82}$$

$$\mathbb{L} = -\frac{1}{2\Omega_\mathfrak{B}} \sum_\alpha \sum_{\beta\neq\alpha} \Big(\boldsymbol{\Phi}^{\alpha\beta}\,\overline{\otimes}\,\big(\mathbf{R}^{\beta\alpha} \otimes \mathbf{R}^{\beta\alpha}\big)\Big), \tag{83}$$

$$^1\mathbb{H} = -\frac{1}{8\Omega_\mathfrak{B}} \sum_\alpha \sum_{\beta\neq\alpha} \big(\boldsymbol{\Phi}^{\alpha\beta}\,\overline{\otimes}\,\big(\mathbf{R}^{\beta\alpha} \otimes \mathbf{R}^{\beta\alpha}\big) \otimes \mathbf{R}^{\beta\alpha}\big), \tag{84}$$

$$^2\mathbb{H} = -\frac{1}{8\Omega_\mathfrak{B}} \sum_\alpha \sum_{\beta\neq\alpha} \big(\boldsymbol{\Phi}^{\alpha\beta}\,\underline{\otimes}\,\big(\mathbf{R}^{\beta\alpha} \otimes \mathbf{R}^{\beta\alpha}\big) \otimes \mathbf{R}^{\beta\alpha}\big), \tag{85}$$

$$\mathbb{M} = -\frac{1}{8\Omega_\mathfrak{B}} \sum_\alpha \sum_{\beta\neq\alpha} \big(\boldsymbol{\Phi}^{\alpha\beta}\,\underline{\otimes}\,\big(\mathbf{R}^{\beta\alpha} \otimes \mathbf{R}^{\beta\alpha}\big) \otimes \mathbf{R}^{\beta\alpha} \otimes \mathbf{R}^{\beta\alpha}\big), \tag{86}$$

$$^1\mathbb{E} = -\frac{1}{24\Omega_\mathfrak{B}} \sum_\alpha \sum_{\beta\neq\alpha} \big(\boldsymbol{\Phi}^{\alpha\beta}\,\overline{\otimes}\,\big(\mathbf{R}^{\beta\alpha} \otimes \mathbf{R}^{\beta\alpha}\big) \otimes \mathbf{R}^{\beta\alpha} \otimes \mathbf{R}^{\beta\alpha}\big), \tag{87}$$

$$^2\mathbb{E} = -\frac{1}{24\Omega_\mathfrak{B}} \sum_\alpha \sum_{\beta\neq\alpha} \Big(\boldsymbol{\Phi}^{\alpha\beta}\,\underline{\underline{\otimes}}\,\big(\mathbf{R}^{\beta\alpha} \otimes \mathbf{R}^{\beta\alpha} \otimes \mathbf{R}^{\beta\alpha}\big) \otimes \mathbf{R}^{\beta\alpha}\Big). \tag{88}$$

According to Eq. (18), (19) and (31), we get

$$\mathbb{H} = -\frac{1}{4\Omega_\mathfrak{B}} \sum_\alpha \sum_{\beta\neq\alpha} \big(\boldsymbol{\Phi}^{\alpha\beta}\,\overline{\otimes}\,\big(\mathbf{R}^{\beta\alpha} \otimes \mathbf{R}^{\beta\alpha}\big) \otimes \mathbf{R}^{\beta\alpha}\big), \tag{89}$$

$$\mathbb{E} = -\frac{1}{12\Omega_\mathfrak{B}} \sum_\alpha \sum_{\beta\neq\alpha} \big(\boldsymbol{\Phi}^{\alpha\beta}\,\overline{\otimes}\,\big(\mathbf{R}^{\beta\alpha} \otimes \mathbf{R}^{\beta\alpha}\big) \otimes \mathbf{R}^{\beta\alpha} \otimes \mathbf{R}^{\beta\alpha}\big), \tag{90}$$

$$\mathbb{W} = -\frac{1}{24\Omega_\mathfrak{B}} \sum_\alpha \sum_{\beta\neq\alpha} \big(\boldsymbol{\Phi}^{\alpha\beta}\,\underline{\otimes}\,\big(\mathbf{R}^{\beta\alpha} \otimes \mathbf{R}^{\beta\alpha}\big) \otimes \mathbf{R}^{\beta\alpha} \otimes \mathbf{R}^{\beta\alpha}\big). \tag{91}$$

From Eq. (86) and (91), it could be found that $\mathbb{M}$ is just three times of $\mathbb{W}$ which leads to no essential difference between the sufficient conditions given by Bardenhagen et. al and us. Detailed expression of $\boldsymbol{\varphi}^{\alpha\beta}$ and $\boldsymbol{\Phi}^{\alpha\beta}$ for EAM potentials are given in Appendix B. Summations represented by $\alpha$ in above expressions run over all atoms in the characteristic volume. With these results, it can be demonstrated that expressions in (53), (54) and (58-62) are the same as the corresponding ones listed above except for $\mathbf{Q}$, $^1\mathbb{H}$, $^2\mathbb{H}$ and $\mathbb{H}$ which are different by a negative sign. Below, we will show that these quantities are zeros. By noting that $(\alpha, \beta)$ in the summation are exchangeable, taking Eq. (89) for example, we have

$$\mathbb{H} = -\frac{1}{4\Omega_{\mathfrak{B}}}\sum_{\alpha}\sum_{\beta\neq\alpha}\left(\boldsymbol{\Phi}^{\alpha\beta}\overline{\otimes}\left(\mathbf{R}^{\beta\alpha}\otimes\mathbf{R}^{\beta\alpha}\right)\otimes\mathbf{R}^{\beta\alpha}\right)$$

$$= -\frac{1}{4\Omega_{\mathfrak{B}}}\sum_{\alpha}\sum_{\beta\neq\alpha}\left(\boldsymbol{\Phi}^{\beta\alpha}\overline{\otimes}\left(\mathbf{R}^{\alpha\beta}\otimes\mathbf{R}^{\alpha\beta}\right)\otimes\mathbf{R}^{\alpha\beta}\right)$$

$$= -\frac{1}{4\Omega_{\mathfrak{B}}}\sum_{\alpha}\sum_{\beta\neq\alpha}\left(\boldsymbol{\Phi}^{\alpha\beta}\overline{\otimes}\left(\mathbf{R}^{\alpha\beta}\otimes\mathbf{R}^{\alpha\beta}\right)\otimes\mathbf{R}^{\alpha\beta}\right).$$

(92)

In the derivation of the last equation above, $\boldsymbol{\Phi}^{\beta\alpha}=\boldsymbol{\Phi}^{\alpha\beta}$ is employed which is valid when the characteristic volume is large enough to avoid effects from surfaces at microscopic scale (but small at macroscopic scale). On the other hand, since $\mathbf{R}^{\beta\alpha}=-\mathbf{R}^{\alpha\beta}$, Eq. (89) is equivalent to

$$\mathbb{H} = \frac{1}{4\Omega_{\mathfrak{B}}}\sum_{\alpha}\sum_{\beta\neq\alpha}\left(\boldsymbol{\Phi}^{\alpha\beta}\overline{\otimes}\left(\mathbf{R}^{\alpha\beta}\otimes\mathbf{R}^{\alpha\beta}\right)\otimes\mathbf{R}^{\alpha\beta}\right). \tag{93}$$

By comparing the two expressions of $\mathbb{H}$ in Eq. (92) and Eq. (93), we have $\mathbb{H}=-\mathbb{H}$ which results in $\mathbb{H}=\mathbf{0}$. Similarly, $^1\mathbb{H}$ and $^2\mathbb{H}$ could also be demonstrated to be zeroes. And $\mathbf{Q}$ is also zero by noting that $\boldsymbol{\varphi}^{\alpha\beta}=-\boldsymbol{\varphi}^{\beta\alpha}$. Different from the results of previous works[16,26], these results are obtained under approximations represented by (78) rather than the restrictions that the crystals should be centro-symmetric. These results suggest that all terms of odd order derivatives of displacements in the expansions of energy vanish at continuum level.

## C. Relationships between the generalized elastic instability criterion and phonon instability criterion

To further understanding the generalized elastic instability criterion established at continuum level, phonon instability criterion is derived under the long wave approximation to the higher order via lattice dynamics approaches. In this section, we use $\ell,\ell'$ to donate lattice indexes and $\alpha,\alpha'$ to donate base atom indexes. Other conventions are the same as previous sections. Under harmonic approximations, equation of motion for lattice atoms is expressed as

$$m_\alpha \ddot{u}_i\binom{\ell}{\alpha} = -\sum_{\ell'\alpha'}\Phi_{ij}\binom{\ell,\ell'}{\alpha,\alpha'}u_j\binom{\ell'}{\alpha'}, \tag{94}$$

where $m_\alpha$ is the mass of the $\alpha$-th base atom. By substituting plane wave solutions of

$$u_j\binom{\ell}{\alpha} = \frac{1}{\sqrt{m_\alpha}}\overline{\Lambda}(\Xi)\hat{e}_j^\alpha(\Xi)\exp\{i[\Xi\mathbf{R}^\ell-\omega(\Xi)t]\}, \tag{95}$$

into Eq. (94), secular equation determining the phonon dispersion relations between the frequency ($\omega$) and wave vector ($\Xi$) is obtained to be

$$\omega^2(\Xi)\hat{e}_i^\alpha = \sum_{\alpha',j}D_{\alpha i,\alpha'j}(\Xi)\hat{e}_j^{\alpha'}, \tag{96}$$

where $\overline{\Lambda}$ and $\hat{e}_i^\alpha$ are the magnitude and vibration direction of the wave, and $\mathbf{D}$ is the dynamical matrix, defined by

$$D_{\alpha i,\alpha'j}(\Xi) = \frac{1}{\sqrt{m_\alpha m_{\alpha'}}}\sum_{\ell,\ell'}\Phi_{ij}\binom{\ell,\ell'}{\alpha,\alpha'}\exp[i\Xi\cdot(\mathbf{R}^{\ell'}-\mathbf{R}^\ell)]. \tag{97}$$

In the remaining of this section, only simple lattices are considered. The obtained results are also applied for multiple lattices by replacing the atom indexes and $\hat{e}_i$ with the lattice indexes and $\hat{e}_i^\alpha$.

Since only the real part of **D** is physically meaningful, from Eq. (96) and (97), we have

$$\frac{1}{m_0}\hat{e}_i\left[\sum_{\alpha,\beta\in\mathfrak{B}}\Phi_{ij}^{\alpha\beta}\cos\left(\Xi_k R_k^{\alpha\beta}\right)\right]\hat{e}_j = \omega^2 \geq 0. \tag{98}$$

At the vicinity of $\Xi = \mathbf{0}$, by using the identity of $\sum_{\alpha,\beta\in\mathfrak{B}}\Phi_{ij}^{\alpha\beta} = 0$ and serial expansions of

$$\cos\left(\Xi_k R_k^{\alpha\beta}\right) = 1 - \frac{\left(\Xi_k R_k^{\alpha\beta}\right)^2}{2!} + \frac{\left(\Xi_k R_k^{\alpha\beta}\right)^4}{4!} - \cdots, \tag{99}$$

inequality (98) becomes

$$\frac{1}{m_0}\hat{e}_i\left\{\sum_{\alpha,\beta\in\mathfrak{B}}\Phi_{ij}^{\alpha\beta}\left[-\frac{1}{2}\left(\Xi_k R_k^{\alpha\beta}\right)^2 + \frac{\left(\Xi_k R_k^{\alpha\beta}\right)^4}{24}\right]\right\}\hat{e}_j$$

$$= \frac{1}{m_0}\hat{e}_i\left\{-\frac{1}{2}\sum_{\alpha,\beta\in\mathfrak{B}}\Phi_{ij}^{\alpha\beta}\Xi_k R_k^{\alpha\beta}\Xi_l R_l^{\alpha\beta} + \frac{1}{24}\sum_{\alpha,\beta\in\mathfrak{B}}\Phi_{ij}^{\alpha\beta}\Xi_k R_k^{\alpha\beta}\Xi_l R_l^{\alpha\beta}\Xi_m R_m^{\alpha\beta}\Xi_n R_n^{\alpha\beta}\right\}\hat{e}_j$$

$$= \frac{1}{m_0}\hat{e}_i\left\{\left(-\frac{1}{2}\sum_{\substack{\alpha\neq\beta\\\alpha,\beta\in\mathfrak{B}}}\Phi_{ij}^{\alpha\beta}R_k^{\alpha\beta}R_l^{\alpha\beta}\right)\Xi_k\Xi_l - \left(-\frac{1}{24}\sum_{\substack{\alpha\neq\beta\\\alpha,\beta\in\mathfrak{B}}}\Phi_{ij}^{\alpha\beta}R_k^{\alpha\beta}R_l^{\alpha\beta}R_m^{\alpha\beta}R_n^{\alpha\beta}\right)\Xi_k\Xi_l\Xi_m\Xi_n\right\}\hat{e}_j$$

$$= \frac{\Omega_\mathfrak{B}}{m_0}\hat{e}_i\{\mathbb{L}_{ikjl}\Xi_k\Xi_l - \mathbb{W}_{ikmjln}\Xi_k\Xi_l\Xi_m\Xi_n\}\hat{e}_j$$

$$= \frac{\Omega_\mathfrak{B}}{m_0}\hat{e}_i\Xi_l\{\mathbb{L}_{ikjl} - \mathbb{W}_{ikmjln}\Xi_m\Xi_n\}\Xi_k\hat{e}_j \geq 0. \tag{100}$$

In the above derivations, Eq. (83) and (91) are used. By noting that $\hat{u}_i^0 = \hat{e}_j$, we find that the sign ahead of $\mathbb{W}$ are opposite in condition (31) and (100), otherwise they are the same. This "sign" paradox has been well-realized in a simple one-dimensional strain-gradient elasticity model[18], which indicates that the requirement of the positive definition for the energy density in the theory of the first strain gradient elasticity is incompatible with the phonon dispersion relations. Several remedies for this dilemma have been proposed phenomenologically[13,17], such as including higher order inertia gradient. Alternatively, by using Eq. (87) instead of Eq. (91), the inequality (100) could be expressed as

$$\hat{e}_i\Xi_l\{\mathbb{L}_{ikjl} - {}^1\mathbb{E}_{ikjlnm}\Xi_m\Xi_n\}\Xi_k\hat{e}_j \geq 0, \tag{101}$$

which is the condition (31) plus an additional term, i.e.,

$$\hat{e}_i\Xi_l\left(\mathbb{M}_{ikljmn} - {}^2\mathbb{E}_{jlnmik}\right)\Xi_m\Xi_n\Xi_k\hat{e}_j = \mathbb{E}_{ikjlnm}\Xi_m\Xi_n\Xi_k\Xi_l\hat{e}_i\hat{e}_j \tag{102}$$

on its left-hand side. This means that the approximate phonon instability criterion (APIC) obtained through serial expansions to the fourth order of $\Xi$ is only a partial higher-order correction of the strain-base instability criteria, while the GEIC is the complete correction up to the fourth order of displacements. Detailed comparisons between these two higher-order instability criteria are performed in the next part of present work.

## IV. Numerical Results for Copper and Aluminum

Traditional strain-based mechanical instability criteria are obtained under the long wave limit assumption, local vibrational modes, represented by short waves, cannot be considered. While GEIC provides an approach to access the roles of these local modes played on the mechanical stabilities of real crystals due to its wavevector dependence. However, due to the sign paradox existing in the first strain gradient elasticity theory, which higher order elastic instability criteria are more appropriate to describe elastic instabilities of crystals remains open. In this part, both GEIC and APIC are employed to predict elastic stabilities of aluminum and copper, while detailed elastic behaviors of these metals under extreme strain rates will be examined through NEMD simulations in the next part. For face-center cubic (FCC) crystals under uniaxial compressions with a compression ratio of $\xi$, the lattice basis vectors of a primitive cell are chosen to be $\mathbf{a}^1 = \frac{a}{2}(0,1,\xi), \mathbf{a}^2 = \frac{a}{2}(1,0,\xi)$ and $\mathbf{a}^3 = \frac{a}{2}(1,1,0)$, where $a$ is lattice constant. Then the corresponding reciprocal lattice vectors are $\mathbf{b}^1 = \frac{2\pi}{a}(-1,1,1/\xi), \mathbf{b}^2 = \frac{2\pi}{a}(1,-1,1/\xi)$ and $\mathbf{b}^3 = \frac{2\pi}{a}(1,1,-1/\xi)$. The lattice constants of copper and aluminum are 3.615 and 4.050 Å, respectively. According to the strain-based instability condition (44), crystals are instable when the minimum eigenvalue ($\lambda_{min}^L$) of $\widetilde{\mathbb{L}}^S$ is negative. As shown in Fig. 1 (a) and (b), the minimum eigenvalue of $\widetilde{\mathbb{L}}^S$ could be determined for copper and aluminum, respectively, as a function of uniaxial compression ratio along [001] direction, where $\widetilde{\mathbb{L}}^S$ is calculated by Eq. (45) and (A.3). EAM potentials developed by Mishin el al [27,28] are employed to describe interatomic interactions of copper and aluminum in present work. Critical strains, defined by the Lagrangian strain at which $\lambda_{min}^L$ begins to be negative, are about -0.022 and -0.027 for aluminum and copper, respectively. From the condition (35) or (32), it is apparent that the GEIC relies not only on elastic constants ($\widetilde{\mathbb{L}}$ and $\widetilde{\mathbb{W}}$) but also on the wavevector. This means that local modes, corresponding to short waves, also contribute to the higher-order elastic instabilities, except for the long waves, represented by strain-related elastic constants. The generalized elastic stability condition (35) is equivalent to require the minimum eigenvalue ($\lambda_{min}^K$) of $\widetilde{\mathbb{K}}$ to be positive for arbitrary $\Xi$ (still a small quantity). Similar statements are also applied for the APIC except for reversing the sign before $\widetilde{\mathbb{W}}$. Comparisons between results obtained by the GEIC and APIC are shown in Fig. 2 where $\lambda_{min}^K$ are calculated at wavevectors with magnitude of 0.01 Å$^{-1}$ and direction along [1 0 0], [1 1 0] and [1 1 1], respectively, for the two criteria in the two metals. It is found that both the aluminum and copper begin to become instable at a strain of about 0.017, slightly smaller than the one predicted by the strain-based instability criterion (See Fig. 1) when $\|\Xi\| = 0.01$ Å$^{-1}$. As a result, instabilities of the deformed crystals could take place before the critical strain predicted by the traditional strain-based criterion is reached. As the increments of the magnitude of $\Xi$, the predicted critical strains by the GEIC and APIC are quite different. The results shown in Fig. 3 and Fig. 4 indicate that the shapes of $\lambda_{min}^K - \xi$ curves predicted by the GEIC among different $\Xi$ are quite similar at relatively large $\|\Xi\|$ for both aluminum and copper. In comparison, the curve shapes predicted by the APIC change dramatically. This means that the GEIC is more numerically stable than the APIC.

Since GEIC, as well as APIC, explicitly depends on $\Xi$ and no prior knowledge could be used to imply the reference wavevector at which crystals begin to become instable, the sufficient condition represented by non-negative minimum eigenvalue of $\mathbb{W}$ is employed to analyze the high-order instabilities of the metals in the remaining of this part. As shown in Fig. 5, the minimum eigenvalue ($\lambda_{min}^W$) of $\mathbb{W}$ are calculated for copper and aluminum under various uniaxial strains. Interestingly, $\lambda_{min}^W$ keeps at a small value nearby zero except for several sharp valleys. And the first valley begins to emerge at a compression ratio of 0.896 for aluminum and 0.887 for copper, whose width is much larger in aluminum than that in copper. It will be seen in the next part that different widths will result in different elastic behaviors under extreme strain rates. In comparison with the sufficient condition of the APIC (See Fig. 6), $\lambda_{min}^W$ tends to decrease with the increment of the compressions except for several small valleys which appear at a compression ratio of about 0.95 for both aluminum and copper. According to the stability condition (28), crystals may be instable only if either $\lambda_{min}^L$ or $\lambda_{min}^W$ is negative. As shown in Fig. 5, overall, the magnitude of $\lambda_{min}^W$ is an order of about 0.1 eV·Å$^{-1}$, which generates a contribute to $\widetilde{\mathbb{K}}$ smaller than 10$^{-3}$ eV·Å$^{-1}$ for $\|\Xi\| < 0.1$. This contribution is relative small compared with $\lambda_{min}^L$ (typically larger than 10$^{-2}$ eV·Å$^{-1}$). However, the contribution will be enlarged to be more than 10 times of the overall value at the valley, which will lead to a notable influence on the elastic instabilities of crystals. This means that elastic instabilities of deformed crystals are dominated by $\lambda_{min}^L$ except for several compression ratios at which $\lambda_{min}^W$ reach its valleys. Later in the next part, it will be noted that this property can be reflected into the elastic responses of crystals under extreme strain rates.

## V. Instabilities of Single Crystalline Copper and Aluminum under Ramp compressions

Strain gradient induced mechanical instabilities are recently observed in iron single crystals under ramp compressions simulated by NEMD simulations[6]. It is found that singularities would arise in wave profile when the instabilities take place in the loaded iron samples. In present work, the ramp compression technique continues to be adopted for studying mechanical instabilities of two typical plastic metals (copper and aluminum). The same interatomic potentials as the ones in Part IV are adopted for copper and aluminum. A copper single crystals, with initial sizes of 18.08×18.08×289.20 nm, are impacted along +Z direction (corresponding to [001] direction) through a moving infinite massive piston at 0K. Ramp compression is generated via linearly increasing the impacting velocity ($v_p$) of the piston from zero to a maximum value ($v_p^{max}$) within a given time ($t_{rising}$). After the ramp compression, the piston keeps its maximum velocity for a certain time ($t_s$) before being removed away from compressed sample. Applied strain rate could be evaluated by $v_p^{max}/(c_L t_{rising})$, where $c_L$ is longitudinal sonic speed along compression direction. For copper and aluminum single crystals, the values of $c_L$ along [001] direction are 3.50 and 5.87 km/s, respectively. Our simulated strain rates range from 10$^9$ to 10$^{10}$ s-1. For aluminum, a single crystal sample, with an initial size of 20.15×20.15×324.00 nm, are employed for the ramp compressions, with $v_{max}$ = 2.0 km/s and $t_{rising}$ = 80 ps, along [001] direction. Besides, lattice deformations of compressed samples are analyzed by a lattice-analyses technique mentioned in ref. [7,29].

Taking the ramp compression on Cu with $v_p^{max} = 2km/s$ and $t_{rising} = 80ps$, and the ramp on Al with $v_p^{max} = 2km/s$ and $t_{rising} = 80ps$ for example, the ranges of the applied strain gradient for the three cases are estimated to be [0.20, 0.32], [0.073, 0.097] and [0.24, 0.28], respectively, where units are 1×10$^{-3}$ Å$^{-1}$. For constantly used EAM potentials, cutoff distance is often less than 1 nm. Under the strain gradients involved our simulations, will lead strain will change less than 10$^{-2}$ within the cutoff distance. Consequently, at any stage of the compressions before instabilities begin, lattice at any position of the simulated crystals is almost under uniform compression albeit with small strain gradient disturbances. This condition is well consistent with the precondition of the instability criteria established in this work. Thus, the theoretical results in Part IV could be employed to explain the elastic instabilities in this case.

Wave profiles represented by particle velocity and strain for copper and aluminum are shown respectively in Fig. 7 and Fig. 8. It is interesting to note that singularities, represented by the knees in the wave profiles, could be observed before plasticity explosively grows. These knees are the origin of "elastic shock waves" under ramp compression. More detailed discussions on the elastic shock waves could be found in Supplementary Materials. For copper, the first and second knees firstly emerge at strains of about -0.042 and -0.114, respectively. For aluminum, the first, second and third knees begin to emerge at about -0.013, -0.032 and -0.104, respectively. Specially, the first major knee for copper and aluminum begins at -0.042 and -0.032. It is interesting that the last knee appears at a strain corresponding to the one where $\lambda_{min}^W$ begins to enter its first valley for both metals. And both the major knees correspond to strains at which $\lambda_{min}^L$ changes its slope (See Fig. 1), while they are slightly smaller than the critical strain predicted by the strain-based instability criterion. The latter one is reasonable since the instable crystals could be carried into a more highly compressed state before transformation into a new stable configuration due to the extreme strain rates. This phenomenon is often termed as "over pressurization". Before the instabilities develop into plastic flows, the over-pressured crystals may enter the valley of the $\lambda_{min}^W - \xi$ curves where contributions from the higher-order terms increase dramatically and thus result in an obvious nonlinear elastic behavior at the corresponding strain states. Thereby, knees after the major one in the wave profile begin to emerge at the strains where valleys of the $\lambda_{min}^W - \xi$ curves are located. Besides, a small knee appears before the major one in the aluminum, while no apparent knees are observed before the major one in copper. This indicates that higher-order instabilities are easier to take place than the strain-controlled ones in aluminum. While, in copper, the higher-order effects are not obvious. To further clarify this point, we use $\varpi = \lambda_{min}^W / \lambda_{min}^L$ to measure the relative contributions of the higher-order terms on the stabilities. For initial aluminum and copper samples, $\varpi$ are 6.45 Å$^2$ and 6.14 Å$^2$, respectively. This means that the relative contributions of the higher-order terms in aluminum are larger than the one in copper so that the high-order contributions are more obvious in aluminum at the early stage of the uniaxial compressions.

## VI. Conclusions and Remarks

In real crystals, contributions from local modes on the mechanical stabilities become unneglectable under severely non-uniform disturbances, such as dynamic loadings. Generalized elastic instability criterion at continuum level is systematically established within frameworks of a higher-order phenomenological theory [8,9]. This instability criterion is a major modification of the high-order elastic instability conditions originally proposed by Bardenhagen et al [11], although the sufficient condition required by stabilities of solids is the same. Our instability criterion has an additional term due to contributions from the third order gradients of displacements which are not considered in the ones by Bardenhagen et al. Later, we find that only including the additional term enable us to reproduce the well-known "sign" paradox in the first strain-gradient elasticity theory. Besides, without considering all contributions of higher-order gradients on strain energies, the established continuum criteria could reproduce well-known results in traditional strain-based theory, such as modified Born criterion and Λ-criterion. And it is found that the established one is equivalent to the Λ-criterion while reduce to the modified Born criterion only under homogenous stresses or pressures. This result indicates that the modified Born criterion is not as precise as the Λ-criterion under heterogenous stress states.

Elastic constants involved in the established instability criteria are evaluated through generalizing the extended Cauchy–Born rule to the fourth-order, whose microscopic expressions are found to be the same as the ones derived via spatial expansions of atom displacements. Because the later approach is only precise enough for crystals binding through short-ranged interatomic interactions (such as crystals binding through metallic or covalent bonding), it could be inferred that the generalized Cauchy–Born rule should also obey this restriction. Considering the nonlinear phonon scattering at zero wavevector up to the fourth order, an approximate phonon instability criterion is established, which are different with the generalized elastic instability criterion by an opposite sign before $\mathbb{W}$ (known as the "sign" paradox). The difference results from the fact that the approximate phonon instability criterion only considers the wave modes near zero (long wave modes) while the generalized elastic instability criterion has not only considered long-wave limit through including the strain elastic constants, but also to some extent considered short-wave modes via including contributions from surroundings represented by the strain gradient (local) terms. Thereby, the "sign" paradox is just an coincidence. Through numerical testing in two real crystals, i.e., single crystalline copper and aluminum, we find that the generalized elastic instability criterion could indeed provide a better overall description on the elastic stabilities than the approximate phonon instability criterion. According our numerical results in copper and aluminum, it could be asserted that contributions from the higher-order terms on strain energies may increase dramatically at some strains and give rise to strong nonlinear elastic behaviors in the compressed crystals. Results from non-equilibrium molecular dynamic simulations on the copper and aluminum confirm this assertion.

# Acknowledgements

This work is supported by the China Postdoctoral Science Foundation (No. 2017M610824), National Natural Science Foundation of China (No. 11772068) and Presidential Foundation of China Academy of Engineering Physics (No. YZJJLX2017011).

Table I. Independent components of the higher-order elastic constants for single crystalline copper at zero strain and temperature, which are calculated using the method of Admal et al and the one in present work, respectively. Units are in eV/Å. EAM potential developed by Mishin el al[27] is employed for both calculations. Components marked by the same special font or symbol are equal in present work.

|  | $\tilde{M}_{111111}$ | $\tilde{M}_{122122}$ | $\tilde{M}_{221221}$ | $\tilde{M}_{111122}$ | $\tilde{M}_{111221}$ | $\tilde{M}_{122221}$ | $\tilde{M}_{122133}$ | $\tilde{M}_{122331}$ | $\tilde{M}_{221331}$ | $\tilde{M}_{123123}$ | $\tilde{M}_{213123}$ |
|---|---|---|---|---|---|---|---|---|---|---|---|
| Admal et al | -0.174 | 0.360 | 0.266 | 0.252 | 0.237 | 0.237 | -0.076 | -0.090 | -0.090 | -0.076 | -0.090 |
| This work | -0.174 | 0.360 | **0.251** | **0.251** | *0.236* | *0.236* | *-0.076* | **-0.090** | **-0.090** | *-0.076* | **-0.090** |

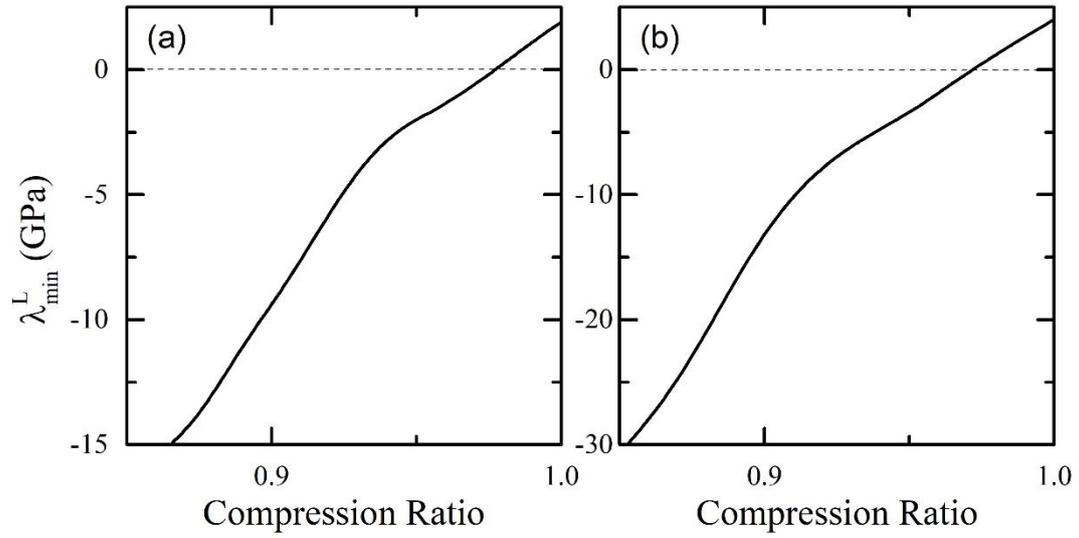

Fig. 1. Minimum eigenvalues of $\widetilde{\mathbb{L}}^S$ as a function of compression ratio for single crystalline (a) aluminum and (b) copper. The results show that $\lambda_{min}^L$ begins to be negative at a compression ratio of 0.9776 and 0.9723 for aluminum and copper, respectively.

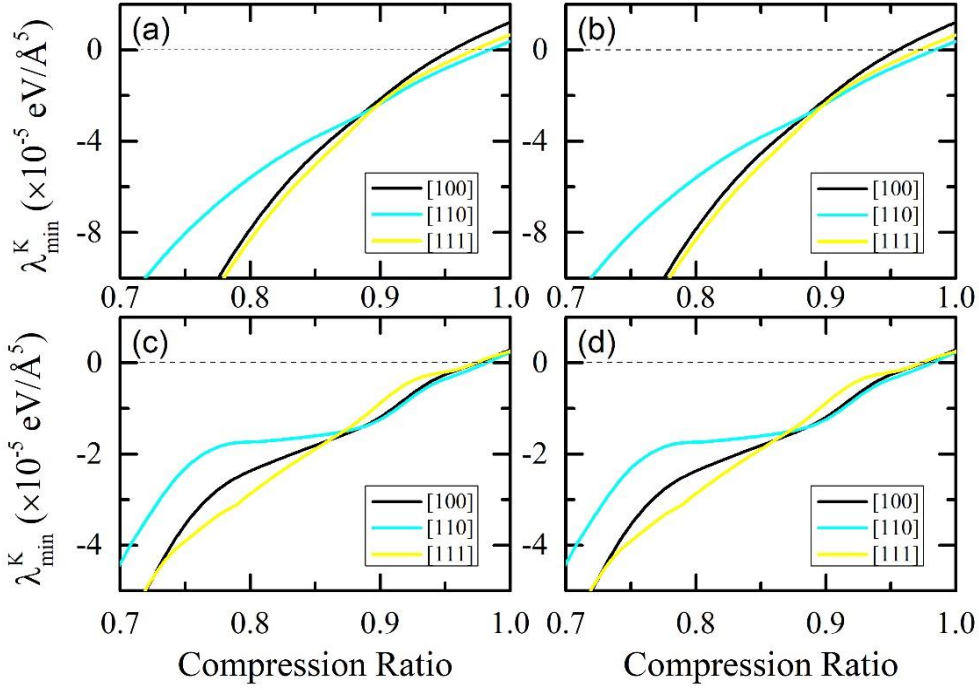

Fig. 2. Minimum eigenvalues of $\widetilde{\mathbb{K}}$ calculated as a function of compression ratio along three wave vectors (pointing along [001], [110] and [111]) with norms of 0.01 Å$^{-1}$ for (a-b) single crystalline copper and (c-d) aluminum, where $\widetilde{\mathbb{K}}_{il}$ is defined by $\left(\mathbb{L}_{iJlM} + \mathbb{W}_{iJKlMN}\Xi_K\Xi_N\right)\Xi_J\Xi_M$ for (a) and (c), and defined by $\left(\mathbb{L}_{iJlM} - \mathbb{W}_{iJKlMN}\Xi_K\Xi_N\right)\Xi_J\Xi_M$ for (b) and (d). The two definitions respectively correspond to the generalized elastic instability criterion and the approximate phonon instability criterion. Similar results are obtained for the two criteria because of the small magnitude of $\Xi$.

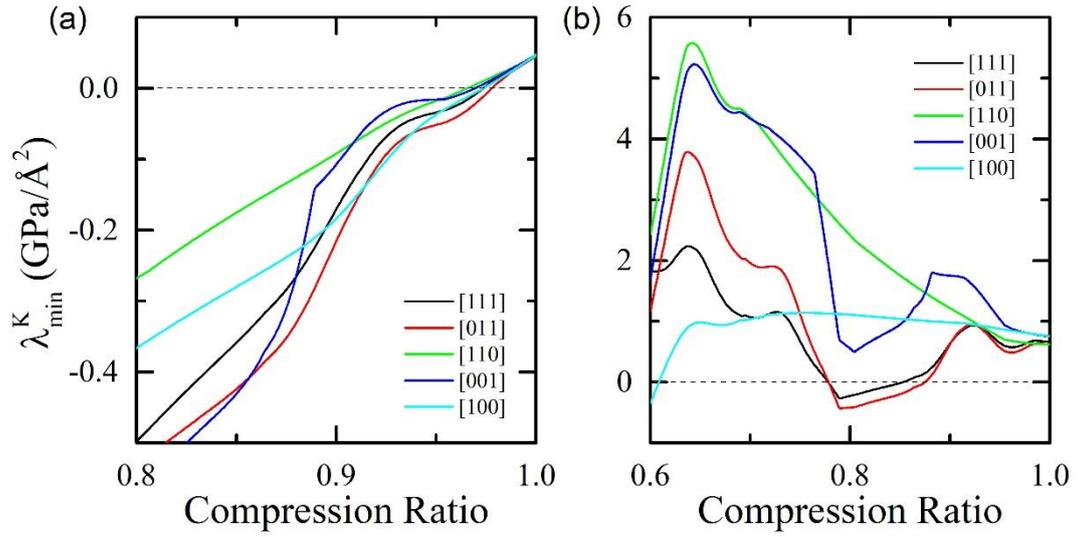

Fig. 3. $\lambda^K_{min}(\Xi)$ calculated as a function of compression ratio for single crystalline aluminum through (a) the generalized elastic instability criterion and (b) the approximate phonon instability criterion, where $\|\Xi\| = 0.1$ Å$^{-1}$.

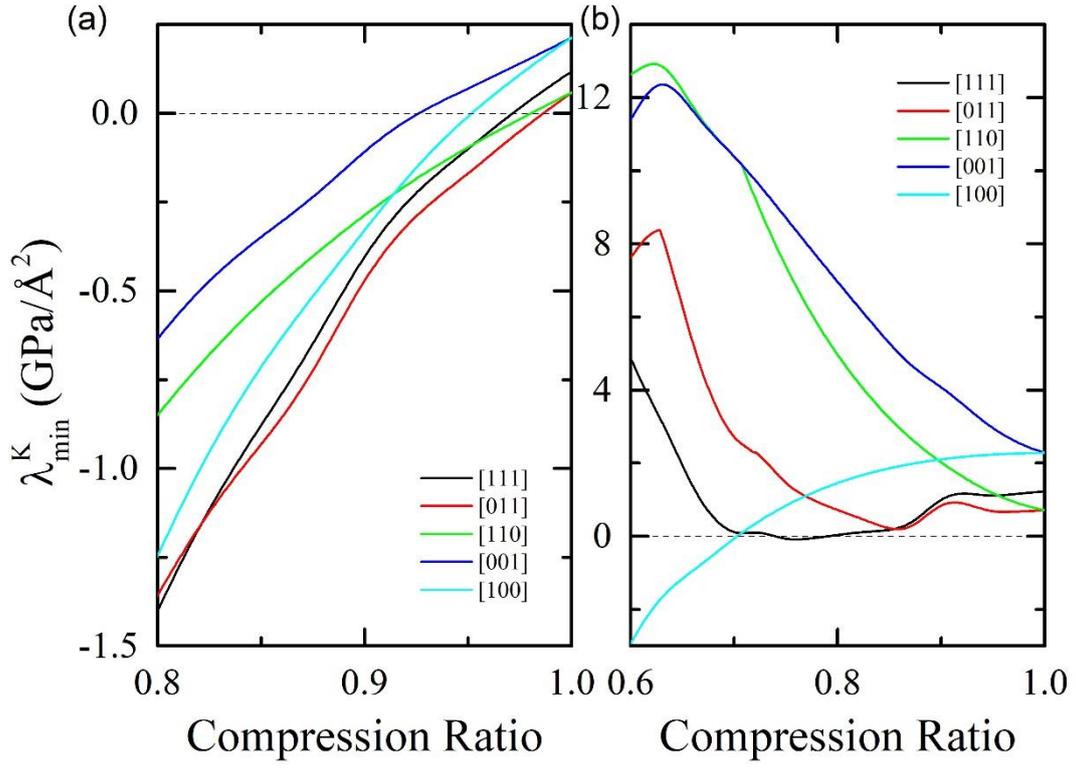

Fig. 4. $\lambda_{min}^{K}(\Xi)$ calculated as a function of compression ratio for single crystalline copper through (a) the generalized elastic instability criterion and (b) the approximate phonon instability criterion, where $\|\Xi\| = 0.1$ Å$^{-1}$.

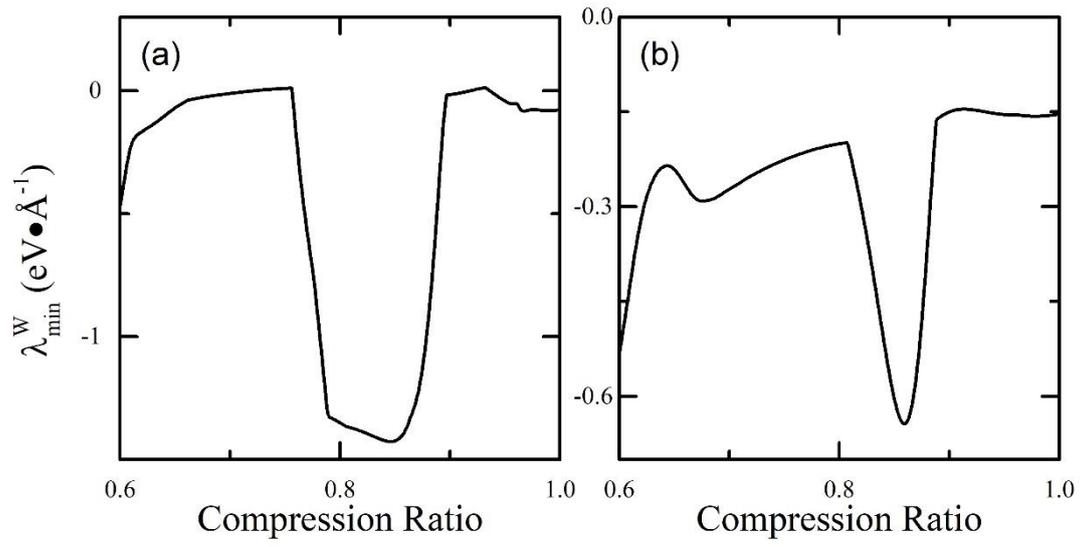

Fig. 5. Minimum eigenvalues of $\widetilde{\mathbb{W}}$ as a function of compression ratio for single crystalline (a) aluminum and (b) copper.

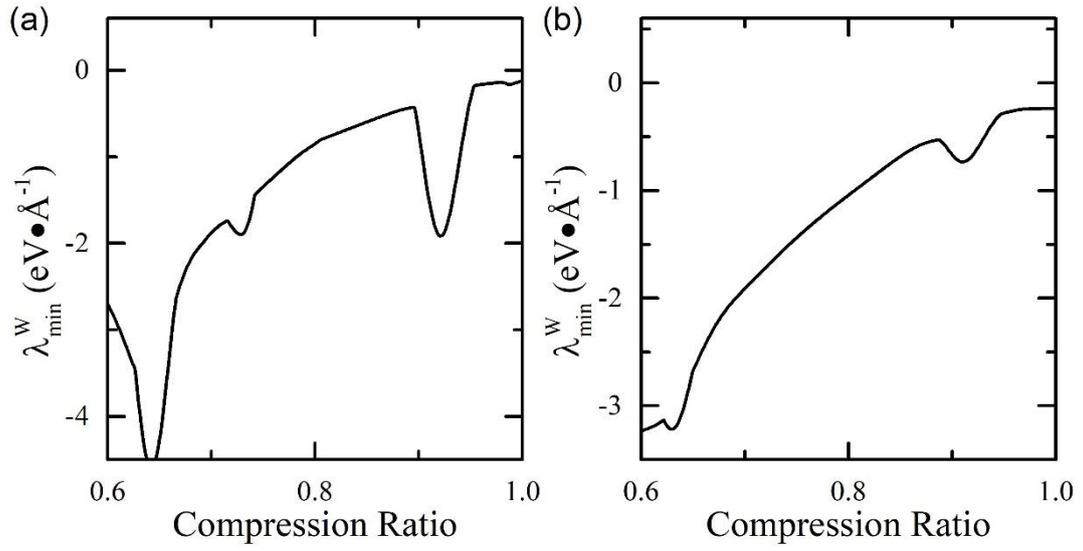

Fig. 6. Minimum eigenvalues of $-\widetilde{\mathbb{W}}$ as a function of compression ratio for single crystalline (a) aluminum and (b) copper.

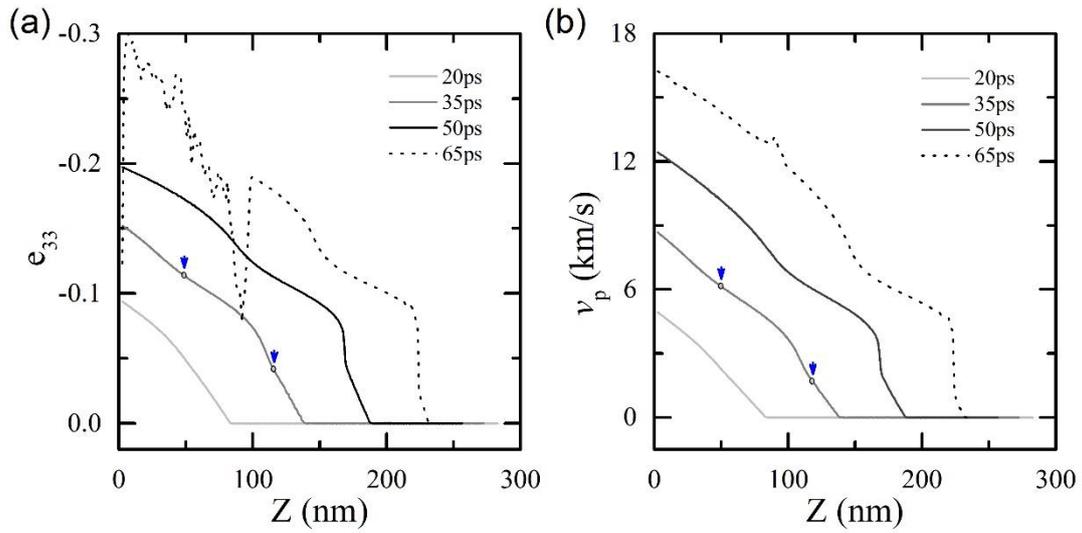

Fig. 7. Wave profile of (a) strain and (b) particle velocity for copper single crystal under ramp compressions, with $v_{max}$ = 2.0 km/s and $t_{rising}$ = 80 ps, along [001] direction. Plasticity takes place in the sample at about 65ps, before which only elastic compressions are observed. At 35ps, initial knees (marked by blue arrows) in the elastic waves are created by mechanical instabilities. And the knee would finally develop into a shock at later time.

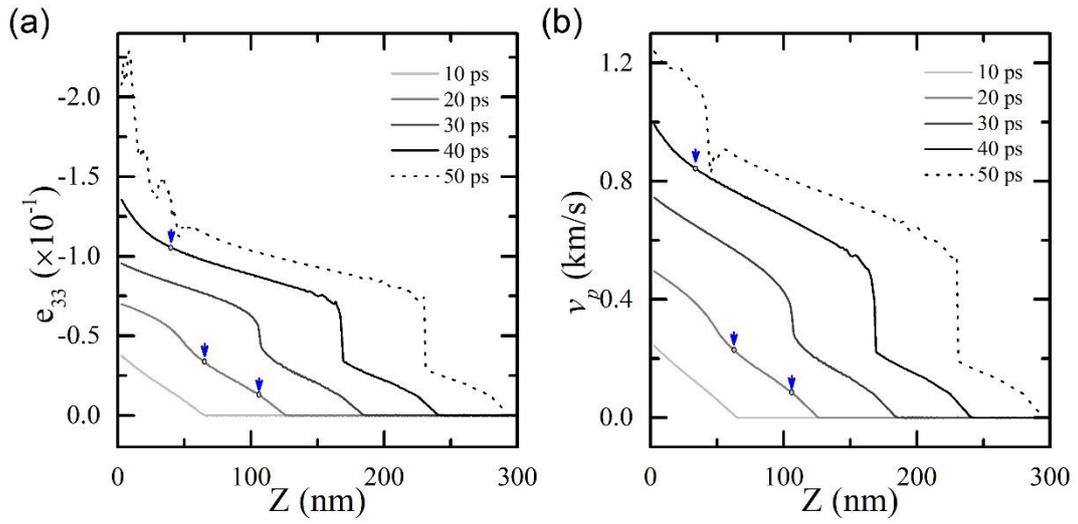

Fig. 8. Wave profile of (a) strain and (b) particle velocity for aluminum single crystal under ramp compression, with $v_{max}$ = 2.0 km/s and $t_{rising}$ = 80 ps, along [001] direction. The aluminum sample is elastically compressed until plasticity takes place at 50ps. Others are the same as Fig. 7.

# Appendix A.

For description convenience, $\mathcal{A}$ is a collection of atoms belonging to one primitive cell whose volume is $\Omega_0$, while $\mathcal{B}$ is a collection of atoms containing all atoms of interests as well as its surroundings. Volume occupied by the atoms of interests is $\Omega_\mathcal{B}$. Typically, $\mathcal{B}$ takes all atoms in simulated systems. Other conventions are the same as the main text. Under small strain gradient disturbances, according to analyses in the main text, $\widetilde{\mathbb{L}}$ and $\widetilde{\mathbb{M}}$ are approximately equal to the ones defined in the current configuration, that is

$$\widetilde{\mathbb{L}}_{iJlM} = \frac{1}{2\Omega_\mathcal{B}} \left\{ \frac{1}{2} \sum_{\alpha \neq \beta} \sum_{\mu \neq \nu} \kappa_{\alpha\beta\mu\nu} \frac{r_i^{\alpha\beta}}{r^{\alpha\beta}} r_J^{\alpha\beta} \frac{r_l^{\mu\nu}}{r^{\mu\nu}} r_M^{\mu\nu} + \sum_{\alpha \neq \beta} \varphi_{\alpha\beta} \frac{1}{r^{\alpha\beta}} \left( \delta_{il} - \frac{r_i^{\alpha\beta} r_l^{\alpha\beta}}{r^{\alpha\beta} r^{\alpha\beta}} \right) r_J^{\alpha\beta} r_M^{\alpha\beta} \right\},$$

(A.1)

$$\widetilde{\mathbb{M}}_{iJKlMN} = \frac{1}{8\Omega_\mathcal{B}} \left\{ \frac{1}{2} \sum_{\alpha \neq \beta} \sum_{\mu \neq \nu} \kappa_{\alpha\beta\mu\nu} \frac{R_i^{\alpha\beta}}{R^{\alpha\beta}} R_J^{\alpha\beta} R_K^{\alpha\beta} \frac{R_l^{\mu\nu}}{R^{\mu\nu}} R_M^{\mu\nu} R_N^{\mu\nu} \right.$$

$$\left. + \sum_{\alpha \neq \beta} \varphi_{\alpha\beta} \frac{1}{R^{\alpha\beta}} \left( \delta_{il} - \frac{R_i^{\alpha\beta} R_l^{\alpha\beta}}{R^{\alpha\beta} R^{\alpha\beta}} \right) R_J^{\alpha\beta} R_K^{\alpha\beta} R_M^{\alpha\beta} R_N^{\alpha\beta} \right\}.$$

(A.2)

If strain is uniform and the characteristic volume is one primitive cell, then the above two expressions reduces to

$$\widetilde{\mathbb{L}}_{iJlM} = \frac{1}{2\Omega_0} \left\{ \frac{1}{2} \sum_{\substack{\alpha \neq \beta \\ \alpha \in \mathcal{A} \\ \beta \in \mathcal{B}}} \sum_{\substack{\mu \neq \nu \\ \mu \in \mathcal{B} \\ \nu \in \mathcal{B}}} \kappa_{\alpha\beta\mu\nu} \frac{r_i^{\alpha\beta}}{r^{\alpha\beta}} r_J^{\alpha\beta} \frac{r_l^{\mu\nu}}{r^{\mu\nu}} r_M^{\mu\nu} + \sum_{\substack{\alpha \neq \beta \\ \alpha \in \mathcal{A} \\ \beta \in \mathcal{B}}} \varphi_{\alpha\beta} \frac{1}{r^{\alpha\beta}} \left( \delta_{il} - \frac{r_i^{\alpha\beta} r_l^{\alpha\beta}}{r^{\alpha\beta} r^{\alpha\beta}} \right) r_J^{\alpha\beta} r_M^{\alpha\beta} \right\},$$

(A.3)

$$\widetilde{\mathbb{M}}_{iJKlMN} = \frac{1}{8\Omega_0} \left\{ \frac{1}{2} \sum_{\substack{\alpha \neq \beta \\ \alpha \in \mathcal{A} \\ \beta \in \mathcal{B}}} \sum_{\substack{\mu \neq \nu \\ \mu \in \mathcal{B} \\ \nu \in \mathcal{B}}} \kappa_{\alpha\beta\mu\nu} \frac{r_i^{\alpha\beta}}{r^{\alpha\beta}} r_J^{\alpha\beta} r_K^{\alpha\beta} \frac{r_l^{\mu\nu}}{r^{\mu\nu}} r_M^{\mu\nu} r_N^{\mu\nu} \right.$$

$$\left. + \sum_{\substack{\alpha \neq \beta \\ \alpha \in \mathcal{A} \\ \beta \in \mathcal{B}}} \varphi_{\alpha\beta} \frac{1}{r^{\alpha\beta}} \left( \delta_{il} - \frac{r_i^{\alpha\beta} r_l^{\alpha\beta}}{r^{\alpha\beta} r^{\alpha\beta}} \right) r_J^{\alpha\beta} r_K^{\alpha\beta} r_M^{\alpha\beta} r_N^{\alpha\beta} \right\}.$$

(A.4)

For EAM potentials, after substituting Eq. (67) and (68) into (A.3) and (A.4), the detailed microscopic expressions of $\widetilde{\mathbb{L}}$ and $\widetilde{\mathbb{M}}$ are obtained as below:

$$\widetilde{\mathbb{L}}_{iJlM} = \frac{1}{2\Omega_0} \sum_{\substack{\alpha \neq \beta \\ \alpha \in \mathcal{A} \\ \beta \in \mathcal{B}}} \left\{ \phi''(r^{\alpha\beta}) \frac{r_l^{\alpha\beta} r_M^{\alpha\beta}}{r^{\alpha\beta}} + F''(\rho_\alpha) f'(r^{\alpha\beta}) \sum_{\substack{\gamma \\ \gamma \neq \alpha}} f'(r^{\gamma\alpha}) \frac{r_l^{\gamma\alpha} r_M^{\gamma\alpha}}{r^{\gamma\alpha}} \right.$$

$$+ F''(\rho_\beta) f'(r^{\alpha\beta}) \sum_{\substack{\gamma \\ \gamma \neq \beta}} f'(r^{\gamma\beta}) \frac{r_l^{\gamma\beta} r_M^{\gamma\beta}}{r^{\gamma\beta}}$$

$$\left. + [F'(\rho_\alpha) + F'(\rho_\beta)] f''(r^{\alpha\beta}) \frac{r_l^{\alpha\beta} r_M^{\alpha\beta}}{r^{\alpha\beta}} \right\} r_i^{\alpha\beta} r_J^{\alpha\beta}$$

$$+ \frac{1}{2\Omega_0} \sum_{\substack{\alpha \neq \beta \\ \alpha \in \mathcal{A} \\ \beta \in \mathcal{B}}} \{\phi'(r^{\alpha\beta}) + [F'(\rho_\alpha) + F'(\rho_\beta)] f'(r^{\alpha\beta})\} \frac{1}{r^{\alpha\beta}} \left( \delta_{il} \right.$$

$$\left. - \frac{r_i^{\alpha\beta} r_l^{\alpha\beta}}{r^{\alpha\beta} r^{\alpha\beta}} \right) r_J^{\alpha\beta} r_M^{\alpha\beta},$$

(A.5)

$$\widetilde{\mathbb{M}}_{iJKlMN} = \frac{1}{2\Omega_0} \sum_{\substack{\alpha \neq \beta \\ \alpha \in \mathcal{A} \\ \beta \in \mathcal{B}}} \left\{ \phi''(r^{\alpha\beta}) \frac{r_l^{\alpha\beta} r_M^{\alpha\beta} r_N^{\alpha\beta}}{r^{\alpha\beta}} + F''(\rho_\alpha) f'(r^{\alpha\beta}) \sum_{\substack{\gamma \\ \gamma \neq \alpha}} f'(r^{\gamma\alpha}) \frac{r_l^{\gamma\alpha} r_M^{\gamma\alpha} r_N^{\gamma\alpha}}{r^{\gamma\alpha}} \right.$$

$$+ F''(\rho_\beta) f'(r^{\alpha\beta}) \sum_{\substack{\gamma \\ \gamma \neq \beta}} f'(r^{\gamma\beta}) \frac{r_l^{\gamma\beta} r_M^{\gamma\beta} r_N^{\gamma\beta}}{r^{\gamma\beta}}$$

$$\left. + [F'(\rho_\alpha) + F'(\rho_\beta)] f''(r^{\alpha\beta}) \frac{r_l^{\alpha\beta} r_M^{\alpha\beta} r_N^{\alpha\beta}}{r^{\alpha\beta}} \right\} r_i^{\alpha\beta} r_J^{\alpha\beta} r_K^{\alpha\beta}$$

$$+ \frac{1}{2\Omega_0} \sum_{\substack{\alpha \neq \beta \\ \alpha \in \mathcal{A} \\ \beta \in \mathcal{B}}} \{\phi'(r^{\alpha\beta}) + [F'(\rho_\alpha) + F'(\rho_\beta)] f'(r^{\alpha\beta})\} \frac{1}{r^{\alpha\beta}} \left( \delta_{il} \right.$$

$$\left. - \frac{r_i^{\alpha\beta} r_l^{\alpha\beta}}{r^{\alpha\beta} r^{\alpha\beta}} \right) r_J^{\alpha\beta} r_K^{\alpha\beta} r_M^{\alpha\beta} r_N^{\alpha\beta}.$$

(A.6)

According to Admal and et al [16], the strain gradient elastic constant $\mathbb{D}$ is defined by

$$\mathbb{D}_{IJKLMN} = \frac{\partial^2 W}{\partial E_{IJ,K} \partial E_{LM,N}},$$

(A.7)

which relates to $\widetilde{\mathbb{M}}$ by

$$\widetilde{\mathbb{M}}_{iJKlMN} = \frac{1}{4} \left( \mathbb{D}_{IJKLMN} + \mathbb{D}_{IJKLNM} + \mathbb{D}_{IKJLMN} + \mathbb{D}_{IKJLNM} \right) \delta_{iI} \delta_{lL}.$$

(A.8)

Matrix representation of the SGE constants are specially investigated for seventeen symmetry class by [30]. They have shown that the higher order elastic constant ($\widetilde{\mathbb{M}}$) could reduce to the following block-diagonal matrix

$$\mathcal{M} = \begin{bmatrix} A_9 & 0 & 0 & 0 \\ & A_9 & 0 & 0 \\ & & A_9 & 0 \\ & & & J_9 \end{bmatrix}, \tag{A.9}$$

where

$$A_9 = \begin{bmatrix} a_{11} & a_{12} & a_{13} & a_{12} & a_{13} \\ & a_{22} & a_{23} & a_{24} & a_{25} \\ & & a_{33} & a_{25} & a_{35} \\ & & & a_{22} & a_{23} \\ & & & & a_{33} \end{bmatrix}, J_9 = \begin{bmatrix} j_{11} & j_{12} & j_{12} \\ & j_{11} & j_{12} \\ & & j_{11} \end{bmatrix}.$$

(A.10)

for cubic lattices under a generalized Viogt notation which contracts a six order tensor ($\widetilde{\mathbb{M}}_{\alpha\mu\nu\beta\lambda\rho}$) into a 18-D matrix ($\mathcal{M}_{mn}$). The three-to-one subscript correspondences ($\nu\alpha\mu \to m$) are listed below:

$111 \mapsto 1$, $221 \mapsto 2$, $122 \mapsto 3$, $331 \mapsto 4$, $133 \mapsto 5$,

$222 \mapsto 6$, $112 \mapsto 7$, $121 \mapsto 8$, $332 \mapsto 9$, $233 \mapsto 10$,

$333 \mapsto 11$, $113 \mapsto 12$, $131 \mapsto 13$, $223 \mapsto 14$, $232 \mapsto 15$,

$123 \mapsto 16$, $132 \mapsto 17$, $231 \mapsto 18$. (A.11)

With the microscopic expressions (A.6), $\widetilde{\mathbb{M}}$ is calculated and rearranged into its Voigt matrix representation for FCC copper. According to Eq. (A.10), independent components of $\mathcal{M}$ are $\mathcal{M}_{1,1}$, $\mathcal{M}_{2,2}$, $\mathcal{M}_{3,3}$, $\mathcal{M}_{1,2}$, $\mathcal{M}_{1,3}$, $\mathcal{M}_{2,3}$, $\mathcal{M}_{2,4}$, $\mathcal{M}_{2,5}$, $\mathcal{M}_{3,5}$, $\mathcal{M}_{18,18}$ and $\mathcal{M}_{17,18}$, which correspond to components of $\widetilde{\mathbb{M}}$ sequentially from left to right hand side in Table I in the main text. Recently, two generalized Cauchy relations are found for $\mathbb{D}$ in cubic metals binding through pair potentials [16], which reduces the number of independence components from 11 to 9. In this work, five generalized Cauchy relations are satisfied by $\widetilde{\mathbb{M}}$, which are $\mathcal{M}_{1,2} = \mathcal{M}_{3,3}$, $\mathcal{M}_{1,3} = \mathcal{M}_{2,3}$, $\mathcal{M}_{2,4} = \mathcal{M}_{18,18}$, and $\mathcal{M}_{2,5} = \mathcal{M}_{3,5} = \mathcal{M}_{17,18}$. From results of Admal and et al (See Table I and Supplementary Materials), only two relations, i.e., $\mathcal{M}_{2,5} = \mathcal{M}_{17,18}$ and $\mathcal{M}_{2,4} = \mathcal{M}_{18,18}$, are always satisfied, while the other three are approximately satisfied. However, all the five relationships are precisely satisfied in present work. Thereby, only six out of the eleven components are independent. It needs to point out that these generalized Cauchy relations remain to be justified by experiments or theories based on calculations from the first principle.